\documentclass[journal]{IEEEtran}
\usepackage{multirow}
\usepackage{xspace}
\newcommand{\MATLAB}{\textsc{Matlab}\xspace}
\usepackage{graphics}
\usepackage{amsmath}
\usepackage{multicol}
\usepackage{siunitx}
\usepackage{amssymb}
\usepackage{mathpazo} 
\usepackage{float}
\usepackage[T1]{fontenc} 
\usepackage{amsmath,amssymb}
\usepackage{subcaption}
\usepackage{mathtools}
\usepackage[capitalise]{cleveref}
\usepackage{epstopdf}
\usepackage{placeins}
\usepackage{mathrsfs}
\usepackage[font=small,labelfont=bf]{caption}
\usepackage{xcolor}	

\ifCLASSINFOpdf
\else
\fi
\begin{document}
%
\title{Autonomous Emergency Collision Avoidance and Stabilisation in Structured Environments}
%
%
%

\author{Shayan Taherian, Shilp Dixit, Umberto Montanaro
        and Saber Fallah
\thanks{S. Taherian, S. Dixit, U. Montanaro and S. Fallah are with the Department
of Mechanical Engineering at University of Surrey, Guildford GU2 7XH, UK  e-mail: ({s.taherian,s.dixit,u.montanaro,s.fallah}@surrey.ac.uk).}
\thanks{Manuscript received April 19, 2005; revised August 26, 2015.}}

%
%

\markboth{Journal of \LaTeX\ Class Files,~Vol.~14, No.~8, August~2015}%
{Shell \MakeLowercase{\textit{et al.}}: Bare Demo of IEEEtran.cls for IEEE Journals}
%



\maketitle

\begin{abstract}

In this paper, a novel closed-loop control framework for autonomous obstacle avoidance on a curve road is presented. The proposed framework provides two main functionalities; (\textit{i}) collision free trajectory planning using MPC and (\textit{ii}) a torque vectoring controller for lateral/yaw stability designed using optimal control concepts. This paper analyzes trajectory planning algorithm using nominal MPC, offset-free MPC and robust MPC, along with separate implementation of torque-vectoring control. Simulation results confirm the strengths of this hierarchical control algorithm which are: (\textit{i}) free from non-convex collision avoidance constraints, (\textit{ii}) to guarantee the convexity while driving on a curve road (\textit{iii}) to guarantee feasibility of the trajectory when the vehicle accelerate or decelerate while performing lateral maneuver, and (\textit{iv}) robust against low friction surface. Moreover, to assess the performance of the control structure under emergency and dynamic environment, the framework is tested under low friction surface and different curvature value. The simulation results show that the proposed collision avoidance system can significantly improve the safety of the vehicle during emergency driving scenarios. In order to stipulate the effectiveness of the proposed collision avoidance system, a high-fidelity IPG carmaker and Simulink co-simulation environment is used to validate the results.   
\end{abstract}


%
\IEEEpeerreviewmaketitle

\section{Introduction}
%
%
%
%
\IEEEPARstart{A}{utonomous} vehicles capable of providing safe and comfortable point-to-point transportation have been an active area of research in academia and automotive industry for the past three decades. In addition to
performing normal driving tasks, the capability of autonomous vehicles to perform emergency manoeuvres
is paramount for their acceptance as a safe and robust mode of end-to-end transportation \cite{Funke2017,Cheng2019}. On detecting the possibility of a collision, an autonomous collision avoidance system needs to perform two main tasks; (\textit{i}) calculate a safe and feasible path to avoid collision and \textit{(ii)} track the calculated path accurately \cite{Xu2019,Boyali2018} while ensuring the lateral-yaw stability \cite{Wang2018a,Khosravani2018} of the vehicle is maintained.\\
Since, collision avoidance forms such a critical part of any autonomous driving feature-set, many different
implementations for such an autonomous system have been proposed in literature. Heuristic searching algorithm such as Diskstra algorithm, $A^*$ and $D^*$ algorithm have been proposed for planning a safe trajectory \cite{Stentz1995,Viking2007}. If a collision free path exists, these search based algorithms can find it. However, a vehicle dynamics and kinematic constraints are not considered while computing the path which can cause safety and feasibility issues for medium and high-speed driving. Incremental search based algorithms like "Rapidly exploring Random Trees"(RRT) have been developed as a trajectory planner \cite{Chen2019}. Although this algorithm includes vehicle kinematic model, the planned trajectory can be jerky, resulting in uncomfortable driving situation. Another method that is widely used among researchers is Model Predictive Control (MPC). In this algorithm the path-planning task is formulated as a finite horizon optimisation problem with vehicle dynamics and collision-avoidance constraints modeled as constraints for the optimisation. However, collision avoidance constraints usually take the form of non-convex constraints. In order to tackle this problem, researchers come across of techniques such as convexification \cite{Franze2016}, changing the reference frame \cite{p8,Karlsson2016} and affine linear collision avoidance constraints \cite{Nilsson2015a}.\\
A successful collision avoidance system does not end with trajectory planning but extends to the capability
of the system to maintain the lateral-yaw stability of the vehicle while performing an evasive manoeuvre
especially under challenging scenarios such as curved road, wet/icy roads, etc. Torque vectoring with its ability
to generate corrective yaw moment to stabilise the planar dynamics of a vehicle provides a proven technique
to improve the stability of a vehicle in limit-handling situations that might be encountered while performing
some of the extreme evasive manoeuvres. As a result, a large body of literature is dedicated for the development
of controller for torque-vectoring, i.e., the control of traction and braking torque of each wheel to generate
a direct yaw moment. Torque-vectoring can increase the cornering response, enhance the agility and guarantee stability in emergency and transient manoeuvre \cite{DeNovellis2015,Lu2016}. In this work, a torque vectoring controller is designed to ensure that the vehicle maintains its stability while performing limit handling evasive manoeuvres even under challenging conditions such as curved roads and/or icy surface.
 \begin{centering}
 	\begin{figure}[t!]
 		\centering
 		\includegraphics[scale=0.9,width=9.0cm]{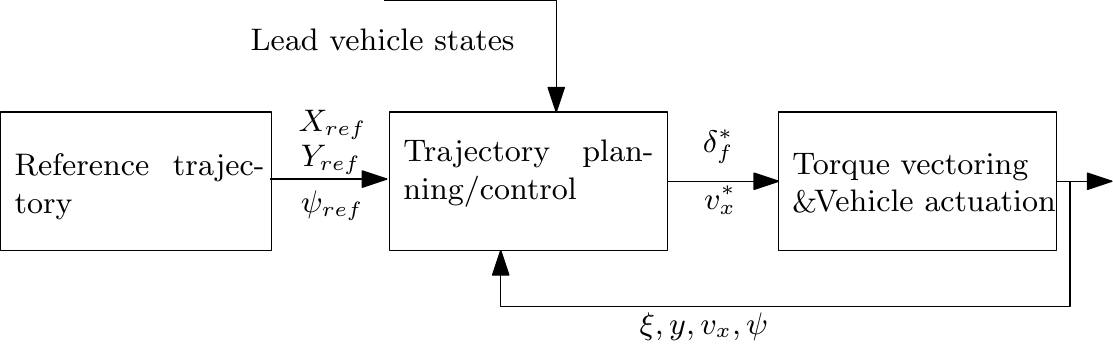}
 		\caption{Control structure}
 		\label{CL_First}
 	\end{figure} 
 \end{centering}
In this paper a closed-loop architecture for collision avoidance and lateral-yaw stabilisation is proposed. The main components of this architecture are (\textit{i}) reference generation block, (\textit{ii}) a trajectory planning/control block, and (\textit{iii}) torque-vectoring control block, see \cref{CL_First}. The reference generation block uses the information from the road/map data to provide the reference for the lane centre. The trajectory generation block based on MPC formulation computes feasible trajectories to safely evade an obstacle in an emergency situation. This reference generation and trajectory planning framework is an extension of our previous work in \cite{Taherian2019} which was applicable only for straight roads. In this paper, this framework has been enhanced to compute safety and collision avoidance constraints for any generic curved road. Furthermore, in adverse conditions (e.g., low friction, wind, etc.) the collision free trajectories generated by the trajectory planner might result in the vehicle operating on the limit of stability. Consequently, to ensure lateral/yaw stability for the subject vehicle while performing the collision avoidance a torque-vectoring controller is used. This controller ensures that the vehicle maintains lateral-yaw stability while performing collision avoidance manoeuvre.\\
The modular structure of the architecture allows to host different implementations of MPC with the same collision avoidance constraints. In this paper, three different MPC techniques (nominal MPC, offset free MPC, and robust MPC) for trajectory planning are tested. The control architeture in \cref{CL_First} is tested in a \MATLAB/IPG-CarMaker co-simulation environment to gain insight on its ability to perform emergency evasive manoeuvres on curved roads. The contribution of this paper are summarised as:  
\begin{itemize}
  \item A mathematical framework to design convex constraints for collision avoidance when travelling on any generic curved road.
  \item A hierarchical motion planning, control and lateral-yaw stability architecture for autonomous collision avoidance in medium and high-speed driving conditions.
  \item Comparison of three different MPC techniques for trajectory planning for collision avoidance and their
impact on the overall performance of the proposed hierarchical control framework, through the definition of the key performance indexes.
\end{itemize}

The paper is organized as follows: \cref{systemmodel} introduces the system models capturing the relevant vehicle and path dynamics for autonomous highway driving that are used for the MPC controllers. \cref{const_Sec} is dedicated to designing collision avoidance constraints along with method of considering convex optimization while driving on a curve. In \cref{Controllers}  three different MPC approaches for trajectory planning are discussed. In \cref{TV}, torque vectoring algorithm is introduced. The effectiveness of the framework to support collision avoidance system is numerically demonstrated in \cref{Results}. Finally, conclusions are presented in \cref{Conclusion}.   
 \begin{centering}
 	\begin{figure}[t!]
 		\centering
 		\includegraphics[scale=0.8,width=6.0cm]{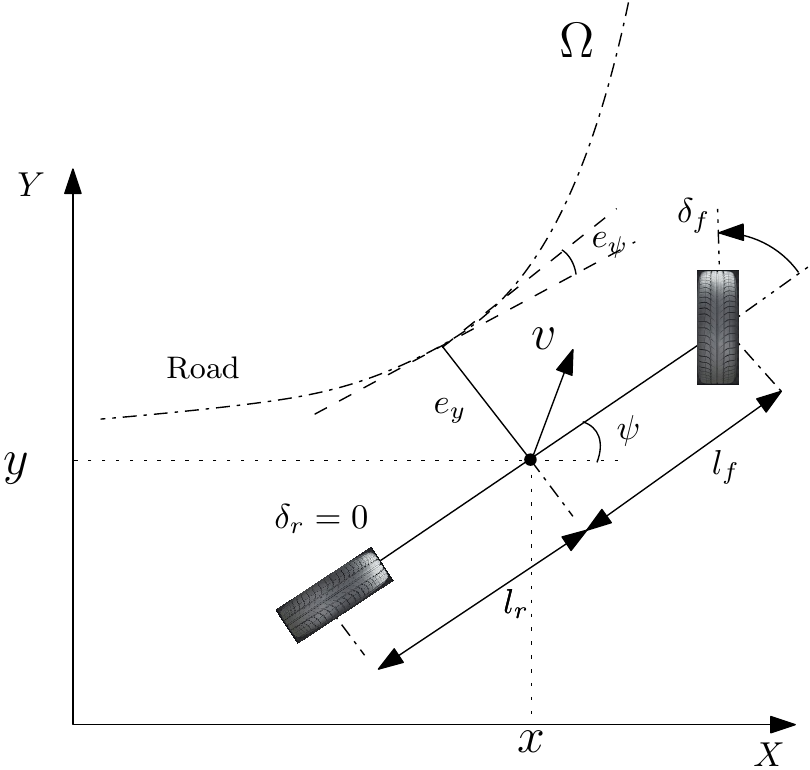}
 		\caption{Kinematic bicycle model}
 		\label{Driving scenario}
 	\end{figure} 
 \end{centering}
\vspace{-0.2cm}
\section{System Model}
\label{systemmodel}
This section provides an overview of the system models that have been used in this paper to describe the vehicle dynamics along with the physical and design constraints. In this paper, two system models are designed and used, (\textit{i}) the single track kinematic vehicle model, which is a common model for trajecory planning algorithm \cite{Kong} in \cref{mod.A}, and (\textit{ii}) a system model that incorporates curvature into dynamics of the system to enhance the tracking performance of the controller \cite{Borrelli2007} in \cref{aug}.
\subsection{Single track vehicle model}
\label{mod.A}
The single track kinematic model (also known as bicycle model) of a vehicle is illustrated in \cref{Driving scenario}. This model captures the planar dynamics of a vehicle in terms of kinematic constraints. Under small angles approximation the system dynamics are expressed as:
 \vspace{-1mm}
\begin{subequations}
\begin{align}
  \dot{x} &= v_x\\
  \dot{y} &= v_x\psi +\frac{l_r}{l_f+l_r}v_x\delta_f\\
  \dot{v}_x &= a_x\\
  \dot{\psi} &= \frac{v_x\delta_f}{l_f+l_r}
\end{align}
\label{system_model}
\end{subequations}
Where $x$ and $y$ are the longitudinal and lateral displacement of the vehicle in inertial coordinates, $\psi$ is the inertial heading angle of the vehicle, $v_x$ is the longitudinal velocity of the vehicle, $l_f$ is the distance of the front axle from centre of gravity (C.G), and $l_r$ is the distance of the rear axle from centre of gravity. The control inputs are steering angle $\delta_f$ and longitudinal acceleration $a_x$. For a given nominal velocity $v_{x,nom}$, the system in \eqref{system_model} can be expressed using a LTI system using the compact notation given below.
  \begin{equation}
      \dot{\xi}=A\xi+Bu
      \label{model}
   \end{equation}
where $\xi\triangleq[x,y,v_{x},\psi]^T\in\mathcal{X}\subseteq\mathbb{R}^4$ is the state vector and $ u\triangleq[\delta_f, a_x]^T\in\mathcal{U}\subseteq\mathbb{R}^2$ is the input vector with $\mathcal{X}$ and $\mathcal{U}$ polyhedron regions depicting state and input constraints respectively. The structure of the system matrix $A$ and input matrix $B$ are as follows:
\begin{equation}
	\begin{bmatrix}
    \dot{x}\\
    \dot{y}\\
    \dot{v}_x\\
    \dot{\psi}
	\end{bmatrix}=
	\underbrace{ 
	\begin{bmatrix}
         0 & 0 & 1 & 0\\
         0 & 0 & 0 & v_{x,\textrm{nom}}\\
         0 & 0 & 0 & 0\\
         0 & 0 & 0 & 0
	\end{bmatrix}}_{A}
	\begin{bmatrix}
	             x\\
                 y\\
                 v_x\\
                 \psi
	\end{bmatrix}+
	\underbrace{ 
	\begin{bmatrix}
                     0 & 0\\
                    \frac{v_{x,\textrm{nom}}l_r}{(l_r+l_f)} & 0\\
                     0 & 1\\
                    \frac{v_{x,\textrm{nom}}}{(l_r+l_f)} & 0
                       \end{bmatrix}
                      }_{B}	
                       \begin{bmatrix}
	                    \delta\\
                        a_x\\
	                    \end{bmatrix}\,
\label{eq:Linear_MPC}
\end{equation}
Where $v_{x,nom}$ is the nominal longitudinal velocity of the vehicle. Then  system \eqref{model} is descretised with a sampling time $t_s$ to obtain linear time invariant discrete system shown below:
   \begin{equation}
      \xi(k+1)=A_d\xi(k)+B_du(k)
      \label{Dis:model}
   \end{equation}
\vspace{-1.2cm}
\subsection{Path dynamics}
 The classical kinematic model presented in \eqref{system_model} formulates the vehicle dynamics in reference to an initial coordinate system. However, it is beneficial to plan the future vehicle motion along the reference curve $\Omega$ in \cref{Driving scenario}. For this reason the classical kinematic model has been modified by introducing new system states $e_y$ as the lateral position error and $e_\psi$ as the heading angle error. In \cref{Driving scenario}, $e_y$ is defined as lateral distance of the centre of gravity of the vehicle from the desired path and $e_\psi$ is defined as the difference between the actual orientation of the vehicle and desired heading angle. These states can be mathematically expressed as \cite{Chatzikomis}:
 \begin{subequations}
 	\begin{align}
 	 e_y &= (y - y_{ref})\cos\psi_{ref}-(x-x_{ref})\sin\psi_{ref}\\
 	    e_{\psi} &= v_x\kappa - \psi
 	\end{align}
\end{subequations}
Where $\kappa$ referred as curvature of the path, $y_{ref}$ is the desired lateral distance, $x_{ref}$ is the desired longitudinal distance and $\psi_{ref}$ is the desired heading angle of the vehicle. Using system \eqref{Dis:model}, the path error dynamics can be expressed as the following discrete LTI system:
\begin{subequations}
	\begin{align}
	\eta(k+1)&=A_{\eta}\eta(k)+B_{\eta}u(k)+B_{\eta,d}d(k)\\
	y(k)&=C_{\eta}\eta(k)+C_{\eta,d}d(k)
	\end{align}
		 \label{off:model} 
	\end{subequations}
Where $ d(k)$ is disturbance in the system. The states and control inputs of this model denoted as $\eta=[x,v_x,y,\psi,e_y,e_{\psi}]^T\in\widetilde{\mathcal{{X}}}\subseteq\mathbb{R}^6$ and $ u\triangleq[\delta_f, a_x]^T\in\widetilde{\mathcal{{U}}}\subseteq\mathbb{R}^2$ respectively, with $\widetilde{\mathcal{X}}$ and $\widetilde{\mathcal{U}}$ being state and input convex constraint sets respectively. 
\subsection{Augmented system}
\label{aug}
Driving on curved roads can reduce the tracking performance of the vehicle due to the curvature of the path which is not captured in the dynamics of the system \eqref{Dis:model}. Hence, incorporating curvature as a disturbance into the dynamics of the system, results in reduction of the tracking error while driving on a curve road. To achieve this, the system model \eqref{off:model} is augmented with the disturbance $d(k)$ to predict the mismatch between the measured and predicted state. The augmented system becomes: 
\begin{equation}
  \begin{gathered}  
	\begin{bmatrix}
    \eta(k+1)\\
    d(k+1)
	\end{bmatrix}
	   =
	   \underbrace{  
	   \begin{bmatrix}
         A_{\eta} & B_{\eta,d}\\
         0 & I
	\end{bmatrix}
	}_{\widetilde{A}}
	 \underbrace{ 
	\begin{bmatrix}
	             \eta(k)\\
                 d(k)
	\end{bmatrix}
}_{\Xi}	  
	  +  
 \underbrace{	  
	  \begin{bmatrix}
                     B_{\eta}\\
                      0
                       \end{bmatrix}
                       }_{\widetilde{B}}	
	                    u(k)\\
	                       \begin{bmatrix}
                            y(k+1)\\
                            d(k+1)
	                         \end{bmatrix}
	                         =
	                    \underbrace{	                         
	                         \begin{bmatrix}
                                      C_{\eta} & 0\\
                                      0 & C_{\eta,d}
	                                       \end{bmatrix}
	                                       }_{\widetilde{C}}	
                                            	\begin{bmatrix}
	                                            \eta(k)\\
                                                 d(k)
	                                          \end{bmatrix}
\label{eq:Offset_MPC}
\end{gathered}
\end{equation}
The system model \eqref{eq:Offset_MPC} is an enhancement of the system \eqref{Dis:model}. The purpose of this transformation is to depict the effectiveness of adding the disturbance into the kinematic vehicle model in collision avoidance manoeuvre.
\section{Cosntraint design}
\label{const_Sec}
 In this section, the mathematical framework for computing convex regions that can be used to plan trajectories on curved roads of varying radii is presented. In the absence of traffic vehicles, the convex region is designed to ensure that the planned vehicle trajectories remain within the boundaries of the road. On the other hand, in the presence of obstacles, the convex region is designed to ensure that the planned trajectories remain within the road boundaries and maintains a safe distance from the obstacle.

\subsection{Road Boundary Constraints}
\begin{centering}
	\begin{figure}[t!]
		\centering
		\includegraphics[scale=0.5,width=6.5cm]{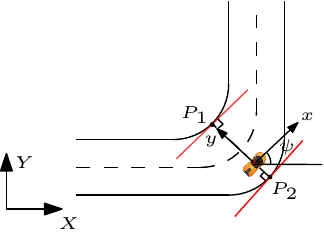}
		\caption {The hyperplanes (red lines) indicate the constraints for optimization. The points on the borders are the projection of the center line of the subject vehicle on the borders}   
		\label{Countor}
	\end{figure} 
\end{centering}

Using the edges of a curved road as constraints within the MPC framework results in non-convex constraints which is not suitable for Quadratic Programming (QP) framework \cite{p12}. Consequently, boundary constraints are designed by assuming that the road segment in the immediate vicinity of the subject vehicle is part of a straight road (\cref{Countor}). The edges of this virtual straight road are obtained where at each time step the task of the MPC controller is to ensure that the planned trajectories lie within the edges of this virtual road.
The point $P_1$ is located at the intersection of the left (outer) road edge with an imaginary line passing through the subject vehicle's centre of gravity (CG) and perpendicular to the subject vehicle's longitudinal axis. The equations of outer lane boundary and intersected line is:
\begin{equation}
\begin{cases}
y_{upper}=f_{upper}(x)\\
ax+by+c=0
\end{cases}
 \label{upper} 
\end{equation} 
The line passing through P1 with a slope $\psi_{ref}$ forms the left (outer) edge of the virtual straight road with equation of:
    \begin{equation}
      m_{upper}x-y+P_{1y}-m_{upper}P_{1x}<0 
   \label{equation upper} 
   \end{equation}   
 Similarly, the line passing through $P_2$ with a slope $\psi_{ref}$ forms the right (inner) edge of the virtual straight road. The inner edge of the virtual straight road can be expressed as:
\begin{equation}
\begin{cases}
y_{lower}=f_{lower}(x)\\
ax+by+c=0
\end{cases} 
 \label{lower} 
\end{equation} 
Consequently in the same way as the upper bound, the lower hyperplane can be defined as:
    \begin{equation}
      m_{lower}x-y+P_{4y}-m_{lower}P_{4x}>0 
   \label{equation lower} 
   \end{equation}   

To decouple the orientation of the virtual straight road with the subject vehicle's orientation, the edges are parameterised using $\psi_{ref}$. It is noted that, this technique is suitable to be implemented for different range of curvature of the road. Moreover, it can be ensured that the optimization problem in MPC will always be feasible, and guarantee the convexity on the curve road. 
\subsection{Collision Avoidance Constraints}
\begin{centering}
	\begin{figure}[t!]
		\centering
		\includegraphics[scale=0.5,width=6cm]{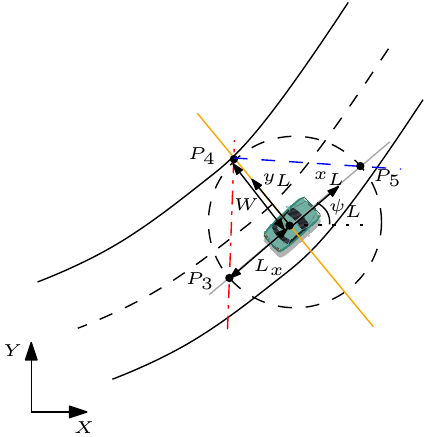}
		\caption{Schematic to construct forward and rear collision avoidance constraints line. \textbf{Note}: $\psi_L$ is lead vehicle orientation, $L_x$ and $W$ indicate the safety distance from the lead vehicle}
		\label{circle-line}
	\end{figure} 
\end{centering}

As mentioned above the collision avoidance constraints are designed to ensure the subject vehicle evades the obstacle vehicle without any collisions. In literature it is common to design ellipsoids/rectangles around the obstacle to obtain the collision avoidance constraints \cite{Gao2014a,Plessen2018}. However, these techniques result in non-convex optimisation problem which is a challenging problem to solve. This paper offers a method to obtain a convex collision avoidance constraints in the form of two affine linear inequalities. The technique is inspired from \cite{p12}, where the collision avoidance constraints are used for only straight driving conditions. In this paper, a general technique has been developed using linear affine collision avoidance constraints that can be implemented for both straight and curve driving conditions. The collision avoidance constraints can be generated using two lines entitled as forward collision avoidance constraints (FCC) and rear collision avoidance constraints (RCC) as shown in \cref{circle-line}. The purpose of generating the FCC is to prevent front end collision while performing a collision free lane change and the purpose of the RCC is to prevent rear end collision while returning to the original lane. The activation of FCC is when the subject vehicle is behind the lead vehicle while RCC is activated when the subject vehicle crosses the lead vehicle. In this framework, three points ($P_3, P_4, P_5$) are required to generate collision avoidance constraints. These points can be calculated by intersecting longitudinal and lateral axis of the lead vehicle (grey and yellow lines respectively) with a circle centred at the lead vehicle's geometrical center, providing the location of the points $P_3$, $P_4$, and $P_5$. The following subsections explain the procedure for generating collision avoidance constraints.
\vspace{0.2cm}
\subsubsection{Forward Collision Avoidance Constraints}

The virtual line on the road segment representing FCC is the line that passes through the points $(P_3,P_4)$ in \cref{FCC}. Point $P_3  (P_{3x},P_{3y})$ is obtained as a results of the intersection of longitudinal axis of the lead vehicle (grey line in \cref{circle-line}):
\begin{equation}
a_{Lon}x_L+b_{Lon}y_L+c_{Lon}=0
 \label{first-point} 
\end{equation}
with a circle centred at the the lead vehicle geometrical centre and radius of $L_x+l_{rL}$:
\begin{equation}
(x-x_L)^2+(y-y_L)^2=(L_x+l_{rL})^2
\end{equation}
 Where $L_x$ is safety distance which can be defined as $L_x=v_xt+L_c$. The longitudinal velocity of the subject vehicle expressed as $v_x$, $t$ is the desired time gap of subject vehicle when approaching lead vehicle and finally $L_c$ is the lead vehicle length. Moreover, Parameters $l_{rL},a_{Lon},b_{Lon},c_{Lon}$ represent the distance of center of gravity of the vehicle to rear wheels and coefficients of the intersected line respectively. It is noted that at every time instant of MPC optimization, $L_x$ updated according to the current value of the subject vehicle velocity ($v_x$). Therefore, depending on the subject vehicle speed, $L_x$ will define the safety distance from the lead vehicle. For generating FCC, another set of points is required $(P_{4x},P_{4y})$. These points are generated by means of intersection of lateral axis of the lead vehicle (yellow line in \cref{circle-line}) with a circle of radius $W$:
\begin{equation} 
(x-x_L)^2+(y-y_L)^2=W^2
\end{equation} 
After defining two set of points, it would be trivial to find the equation of the line for forward collision avoidance constraints, where the final equation will be:
    \begin{equation}
     	\underbrace{m_{FCC}}_{a_{FCC}}x+(\underbrace{-1}_{b_{FCC}})y+\underbrace{P_{1y}-m_{FCC}P_{1x}}_{c_{FCC}}>0 
   \label{equation FCC} 
   \end{equation} 
Where $m_{FCC}$ is the slope of the forward collision avoidance constraint.  The generated FCC divides the $(x,y)$ plane into two regions. Region 1: $a_{FCC}x+b_{FCC}y+c_{FCC} > 0$ which is the safe region, Region 2: $a_{FCC}x+b_{FCC}y+c_{FCC}<0$ which represent the unsafe region. FCC forces the vehicle to be in a safe region while performing a lane change on a curve/straight road. Moreover, \eqref{equation FCC} represents a linear affine constraint approximation which can be formulated in QP format.
\subsubsection{Rear Collision Avoidance Constraints}
\begin{centering}
 	\begin{figure}[t!]
 		\centering
 		\includegraphics[scale=0.5,width=5.8cm]{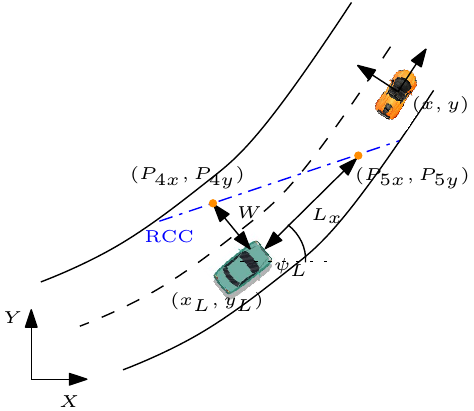}
 		\caption{Schematic to calculate RCC. \textbf{Note}: orange car is subject vehicle, green car is lead vehicle, $L_x$ and $W$ indicate the safety distance from the lead vehicle}
 		\label{RCC}
 	\end{figure} 
 \end{centering}

 Similar to FCC, RCC can be generated with the same structure. The procedure for calculating the intersection point $P_3$, is identical to FCC, thus resulting in generating the equation of the RCC line (\cref{RCC}): 
      \begin{centering}
 	\begin{figure}[t!]
 		\centering
 		\includegraphics[scale=0.5,width=5.5cm]{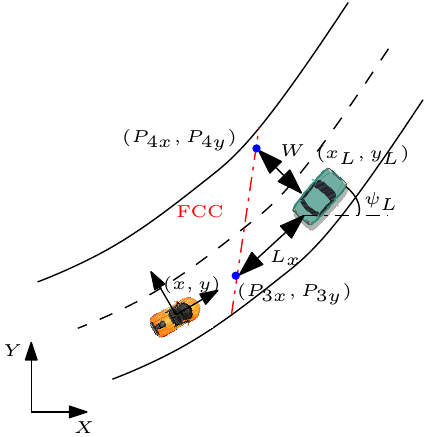}
 		\caption{Schematic to calculate the FCC. \textbf{Note}: orange car-subject vehicle, green car-lead vehicle, red line RCC, $L_x$ and $W$ indicate the safety distance from the lead vehicle}
 		\label{FCC}
 	\end{figure} 
 \end{centering}
    \begin{equation}
     \underbrace{m_{RCC}}_{a_{RCC}}x+(\underbrace{-1}_{b_{RCC}})y+\underbrace{P_{3y}-m_{RCC}P_{3x}}_{c_{RCC}}<0 
   \label{equation RCC} 
   \end{equation}   
It is noteworthy that the aforementioned collision avoidance constraints can be adopted for the scenarios where $(i)$ the subject vehicle needs to change the lane while performing the collision avoidance manoeuvre and $(ii)$ where more traffic member on the road are presented  which require multiple hyperplanes.
\\
\vspace{-0.3cm}
\textit{\textbf{Remarks}}\\
\begin{itemize}
  \item One of the benefits of the proposed approach lies into the fact that there is only one parameter to be tuned (desired time gap) and the rest of the parameters are based on the geometry of the lead vehicle.
   \item The design of the collision avoidance constrains is based on basic mathematical operation, therefore the aforementioned constraints for each traffic vehicle can be generated without any major computation overhead, thus it is suitable for real time implementation. 
\end{itemize}
It is noted that, by using boundary constraints and the appropriate FCC/RCC based constraints a convex region representing safe zones on a curved road can be computed. When applied to an MPC formulation, these constraints guarantee that the planned trajectories are collision free.

\section{Trajectory Planning Controllers}
\label{Controllers}
In general, autonomous highway driving involves motion planning and control of a subject vehicle in order to maintain the velocity and keeping the vehicle away from lane boundaries while avoiding possible collisions. The choice of manoeuvre to perform is the results of the decision making process with the objective of following the desired trajectory $x_{ref}, y_{ref}$ (\cref{reference}) while respecting physical and design limitations of the subject vehicle. In this section, using the definition of the linear prediction model in \cref{systemmodel} and linear system constraints in \cref{const_Sec}, a constrained linear quadratic optimal control problem for three different MPC's is formulated over a prediction horizon $N$. 
\vspace{-0.2cm}
\subsection{Reference Trajectory Generator}
\label{reference}
The reference trajectory generator provides longitudinal and lateral positions ($x_{ref}(s), y_{ref}(s)$) as well as orientation ($\psi_{ref}(s)$) of the road for the vehicle to follow. The reference trajectories are formulated using clothoids method which expressed by Fresnel integrals \cite{Scheuer1997,Shin1992}. The trajectories are parametrized by curvature $\kappa$ as a function of distance $s$ along the path as follows:  
 \begin{equation}
  \psi_{ref}(s)= \int_{0}^{s}\kappa(s)dx
 \label{Clothoid_heading}
 \end{equation}
  \begin{equation}
  x_{ref}(s)= \int_{0}^{s}\cos\psi_{ref}(x)dx
 \label{Clothoid_X}
 \end{equation}
  \begin{equation}
  y_{ref}(s)= \int_{0}^{s}\sin\psi_{ref}(x)dx
 \label{Clothoid_Y}
 \end{equation}
\subsection{Trajectory Planning: Nominal MPC}
For the trajectory planning using nominal linear MPC, System \eqref{eq:Linear_MPC} is subjected to the following state and input constraints: 
  \begin{equation}
      \xi\in\mathcal{X}, u\in\mathcal{U} 
      \label{sys:constr}
  \end{equation}
where $\mathcal{X}=\{\xi\in\mathbb{R}^4:\xi_{min}$$\leq$$\xi$$\leq$$\xi_{max}\}\subset\mathbb{R}^4$ and $\mathcal{U}=\{u\in\mathbb{R}^2:u_{min}$$\leq$$u$$\leq$$u_{max}\}\subset\mathbb{R}^2$  are states and inputs polytope admissible regions (subscripts $min$ and $max$ are the minimum and maximum of the corresponding values). The following cost function is formulated:
  \begin{equation}
  \begin{gathered}
   \mathcal{J}= \sum_{k=0}^{N-1}[\parallel \xi_{ref}-\xi_{t+k\mid t}\parallel_{Q}^2+\parallel u_{t+k\mid t}\parallel_{R}^2]\\
   +\parallel \xi_{ref}-\xi_{t+N\mid t}\parallel_{P}^2
      \end{gathered}
      \label{cost:Nom}
\end{equation}
where $\xi_{t+k|t}$ is the predicted state trajectory at time $t+k$ obtained by applying the control sequence ${U}_t=[{u_t^T,\dots,u_{t+N-1}^T}]^T$ to the system \eqref{Dis:model}, starting from initial state of $\xi_{t|t}$. The parameter $N\in\mathbb{N}^+$is the prediction horizon while $Q\in\mathbb{R}^{4\times 4}$, $P\in\mathbb{R}^{4\times 4}$ and $R\in\mathbb{R}^{2\times 2}$ are weighting matrices. The performance index in \eqref{cost:Nom} consists of the stage cost, input cost and the terminal cost, respectively. The desired state $\xi_{ref}$ representing the reference state for the subject vehicle and is defined as $\xi_{ref}=[x_{ref}, y_{ref}, v_{xref}, \psi_{ref}]^T$. It is noted that in this paper $v_{xref}$ is taken as the initial value of the subject vehicle's velocity. The following constrained optimization problem, for each sampling time, is formulated as:
 \begin{subequations}
	\begin{align} 
	\underset{U_{t}}{\text{min}} \mathcal{J}(u_{t};\xi(t),\xi_{ref})
	\label{prob}
	\end{align}
	\vspace{0.00001cm}
	\hspace*{0.8in}
subject to	
	\vspace{0.00001cm}
	\begin{align}
   \eqref{Dis:model},\eqref{equation upper},\eqref{equation lower},\eqref{equation FCC},\eqref{equation RCC},\eqref{sys:constr}
   \label{const:prob}
\end{align}
\end{subequations}
In this framework, at every time step, the problem \eqref{prob}, under the constraints \eqref{const:prob} are calculated based on the current state $\xi(t)$, over a shifted time horizon. The control input is calculated as  $u(x(t))=U_{t}^{*}(0)$, which is a solution to the problem \eqref{prob}. As the sets $\mathcal{X}$ and $\mathcal{U}$ are convex, then the MPC problem \eqref{prob} is solved as a standard QP \cite{p12} optimisation problem.

\subsection{Trajectory Planning: Offset-free MPC}

For trajectory planning using offset-free MPC system model \eqref{eq:Offset_MPC} is used. This system is subjected to the following state and input constraints:
 \begin{equation}
      \eta\in\widetilde{\mathcal{X}}, u\in\widetilde{\mathcal{U}}
      \label{off:const} 
  \end{equation}
where $\widetilde{\mathcal{X}}=\{\eta\in\mathbb{R}^6:\eta_{min}$$\leq$$\eta$$\leq$$\eta_{max}\}\subset\mathbb{R}^6$ and $\widetilde{\mathcal{U}}=\{u\in\mathbb{R}^2:u_{min}$$\leq$$u$$\leq$$u_{max}\}\subset\mathbb{R}^2$  are states and inputs polytope.
According to this model the following cost function can be written: 
\begin{equation}
\begin{gathered}
\mathcal{J}=  \sum_{k=0}^{N-1}[\parallel \Xi_{ref}-\Xi_{t+k\mid t}\parallel_{Q_{\Xi}}^2+\parallel u_{t+k\mid t}\parallel_{R}^2]\\
+\parallel \Xi_{ref}-\Xi_{t+N\mid t}\parallel_{P_{\Xi}}^2
\end{gathered}
\label{cost:off}
\end{equation}
At each sampling time, the following constrained finite-time control problem is solved:
 \begin{subequations}
	\begin{align} 
	\underset{U_{t}}{\text{min}} \mathcal{J}(u_{t};\Xi(t),\Xi_{ref})
	\label{prob:off}
	\end{align}
\vspace{0.00001cm}
 \hspace*{0.8in}
  subject to
  	\begin{align}
  \eqref{eq:Offset_MPC},\eqref{equation upper},\eqref{equation lower},\eqref{equation FCC},\eqref{equation RCC}, \eqref{off:const}
  \end{align}
\end{subequations}
\textit{\textbf{Remarks}}\\
\vspace{-0.2cm}
\begin{itemize}
  \item $\eta$ and $d$ are parameters while $u$ is the decision variable of the optimisation \eqref{prob:off}.
  \item The terms of the cost function under the summation in \eqref{cost:off}, penalize the state for deviation of reference state as well as penalty on control input, and the last term is penalty on terminal state. 
  \end{itemize}
\subsection{Trajectory Planning: Robust MPC}
\label{RMPC}
This section provides an outline of the robust tube based MPC approach provided in \cite{Mayne2010}. Robust MPC concerns control of the systems that are uncertain, such that the predicted behaviour based on the nominal system is not identical to the actual one. The dynamics of the lateral and yaw motion of the vehicle with its longitudinal velocity have a non-linear relationship. This non-linear relation between the states can be expressed as linear additive disturbance as in \cite{Dixit2019}. The uncertainty in this formulation is considered as parametric uncertainty in linear constraint system. System dynamics in \eqref{eq:Linear_MPC} are rewritten as a linear time invariant system subject to an additive disturbance and can be recast as a linear parameter varying (LPV) system
      \begin{equation}
     \dot{\xi}_p = A_p(v)\xi_p + B_p(v)u
     \label{RMPC-model} 
      \end{equation}
System \eqref{RMPC-model} can be discretised to generate a LPV discrete system:
          \begin{equation}
     {\xi}_p(k+1) = A_d(v)\xi_p(k) + B_d(v)u(k)
     \label{RMPC-dmodel}
      \end{equation}
In which the parameter $p=(A_d(v),B_d(v))$ can take any value in the convex set $\mathcal{P}$ defined by 
             \begin{equation}
               \mathcal{P}=co\{(A_{d,j}(v_j),B_{d,j}(v_j))| j\in\mathscr{J}\}
            \label{Pset}
            \end{equation}
Where $\mathscr{J}=\{1,2,...,J\}$. The system is subject to the bounded convex sets:
                  \begin{equation}
                  \xi\in\mathcal{X}\subset{R^4},  u\in\mathcal{U}\subset{R^2}
                  \end{equation}
With $\mathcal{X}$ and $\mathcal{U}$ are assumed to be compact and polytopic and contains the origin in their interior. According to the dynamics of LPV system \eqref{RMPC-dmodel}, the  nominal system:
                  \begin{equation}
                   \bar{\xi}(k+1) = \bar{A}\bar{\xi}(k) + \bar{B}\bar{u}(k)
                   \label{RMPC-model_nom} 
                     \end{equation}
in which the pair ($\bar{A},\bar{B}$) is obtained by the terms given below \cite{Mayne2010}:
                  \begin{equation}
                  \bar{A}=(\frac{1}{J})\sum_{j=1}^{J} A_{dj}(v_j), \bar{B}=(\frac{1}{J})\sum_{j=1}^{J} B_{dj}(v_j)     
                  \end{equation}      
The system difference equation \eqref{RMPC-dmodel} can be expressed as: 
                           \begin{equation}
                         \xi(k+1) = \bar{A}\xi(k)+\bar{B}u(k)+w(k)
                         \label{uncertain-system}
                         \end{equation} 
Where the disturbance $w$ is defined as 
                  \begin{equation}
                   w =(A(v)-\bar{A})\xi+(B(v)-\bar{B})u
                   \label{w} 
                  \end{equation}                                          
thus the disturbance $w$ is bounded by the set $\mathcal{W}$ defined as:
                  \begin{equation}
                    \begin{gathered}  
          \mathcal{W} =\{(A_d(v)-\bar{A})\xi(k)+(B_d(k)-\bar{B})u(k)|\\(A_d(v),B_d(v))\in\mathcal{P},(\xi,u)\in\mathcal{X}\times\mathcal{U}\} 
                   \end{gathered}
                  \end{equation}                  
The set constraints for the nominal model are selected such that if the closed-loop solution for the nominal system satisfies $(\bar{\xi}(k),\bar{u}(k))\in\bar{\mathcal{X}}\times\bar{\mathcal{U}},\forall{k}$, then $(\xi(k),u(k))\in\mathcal{X}\times\mathcal{U}$ ,where $\bar{\mathcal{X}}$ and $\bar{\mathcal{U}}$ are tightened states and inputs constraints used for robust MPC. Essentially the idea is to steer the uncertain system \eqref{uncertain-system} towards a given reference state, using MPC approach with modified state and input constraints ($\bar{\mathcal{X}},\bar{\mathcal{U}}$). The tightened constraints for the nominal model can be expressed as: 
                        \begin{equation}
                   \bar{\mathcal{X}}=\mathcal{X}\ominus\mathcal{Z},\, \bar{\mathcal{U}}=\mathcal{U}\ominus K\mathcal{Z},
                   \label{const:Robsut}
                        \end{equation}  
Where $K\in{\mathbb{R}^{2\times4}}$ such that $A_k=(\bar{A}+\bar{B}K)$ is hurwitz, and $\mathcal{Z}$ is robust positively invariant set \cite{Rakovic2005} for the system $e(k+1)=A_ke(k)+w(k)$ with $e=(\xi-\bar{\xi})$, such that
                        \begin{equation}
                        A_k\mathcal{Z}\oplus\mathcal{W}\subseteq\mathcal{Z}
                        \end{equation} 
In \cite{Mayne2010} it is proven that if $\bar{\mathcal{X}}$ and $\bar{\mathcal{U}}$ are non-empty sets, they contain the set-points and control inputs that can be robustly imposed to the system \eqref{uncertain-system} when $e(0) = \xi(0)-z(0)$, under the control action:
                 \begin{equation}
                   u = \bar{u}+Ke, \,\bar{u}\in\bar{\mathcal{U}}
                   \label{RMPC-controlaction} 
                     \end{equation}           
These definitions above can be used to formulate the optimisation framework for a tube-based MPC problem \cite{Gao2014}. The cost function can be written as :
  \begin{equation}
\begin{gathered}
\mathcal{J}= \sum_{k=0}^{N-1}[\parallel \xi_{ref}-\bar{\xi}_{t+k\mid t}\parallel_{Q}^2+\parallel \bar{u}_{t+k\mid t}\parallel_{R}^2]\\
+\parallel \xi_{ref}-\bar{\xi}_{t+N\mid t}\parallel_{P}^2
\end{gathered}
\label{cost}
\end{equation}
Where $\bar{\xi}_{t+k\mid t}$ denotes the predicted state at time $t+k$ calculated by applying control sequence $\bar{U}_t=\{\bar{u}_{t,t}...,\bar{u}_{t+k,t}\}$ to the system \eqref{RMPC-dmodel} with $\bar{\xi}_{t,t}=\bar{\xi}(t)$. The matrices $Q$,$R$ and $P$ are weights of appropriate dimension penalising the state tracking error, control action and terminal state. The optimisation problem is defined as:
 \begin{subequations}
	\begin{align} 
  \begin{gathered}
	\underset{\bar{U}_{t}}{\text{min}} \mathcal{J}(\bar{u}_{t};\bar{\xi}(t),\xi_{ref})
      \end{gathered}
      \label{cost_Robust}
\end{align}

	\vspace{0.00001cm} 
	\hspace*{0.8in}
subject to  
	\begin{align}
    \eqref{equation lower},\eqref{equation FCC},\eqref{equation RCC}, \eqref{RMPC-model_nom},\eqref{const:Robsut} 
    \label{opt:Robust} 
\end{align}
\end{subequations}
      \begin{center}
 	\begin{figure*}[t!]
   	\begin{multicols}{2}
   	\includegraphics[width=2\linewidth]{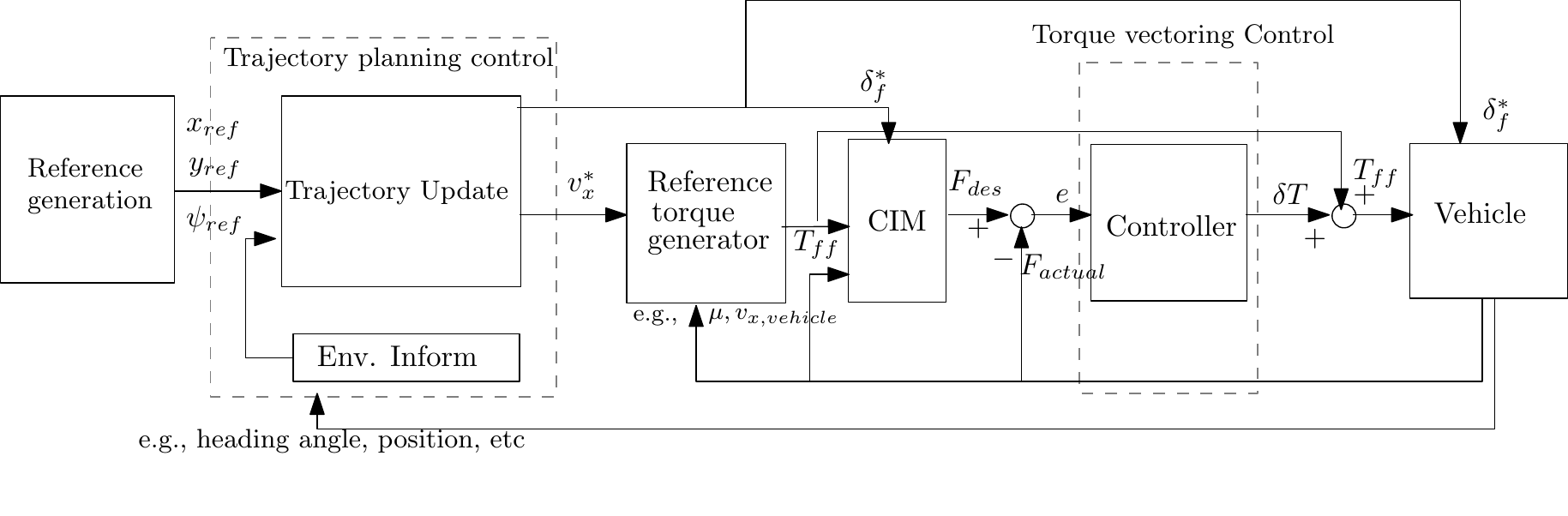}
   	\end{multicols}
   	\vspace*{-5mm}
   	\caption{Closed-loop framework for collision avoidance system}
   	   \label{controlsystem}
   \end{figure*}
\end{center} 
In QP framework, at each time instant $k$ the optimization problems \eqref{prob}, \eqref{prob:off} and \eqref{cost_Robust} are solved subject to the physical and design constraints of the vehicle as well as collision avoidance constraints \eqref{equation FCC},\eqref{equation RCC},\eqref{equation upper} and \eqref{equation lower}. This will lead to obtain the optimal trajectory $\xi^*=[x^*,y^*,v_x^*,\psi^*]$ by simulating the vehicle model \cref{Driving scenario} with optimal inputs $u^*=[\delta^*, a_x^*]$ from the solution of the MPC problems. The optimal trajectories are then passed to the torque vectoring controller (lower-level control) presented in next session as reference signals. The closed-loop structure shown in \cref{controlsystem} summarizes all the steps to perform a collision avoidance in this paper.

\section{Torque vectoring}
\label{TV}
\begin{centering}
	\begin{figure}[tbhp!]
		\centering
		\includegraphics[width=0.6\columnwidth]{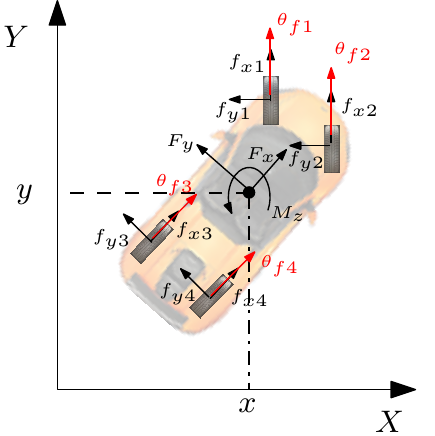}
		\caption{Force conventions}
		\label{Force_conv}
	\end{figure}
\end{centering}

A torque-vectoring controller was developed to maintain lateral/yaw stability of the vehicle even in limit-handling situations. A proportional controller is used to calculate the braking/traction torque required to eliminate the error between optimal velocity from the MPC and actual velocity of the vehicle \cite{Jalali2012}. The following equation represents the required torque which is equally distributed among the wheels: 
    \begin{equation}
    T_{ff}=K_p(v_{x}^*-v_x) 
   \end{equation}
where $K_p$ is the proportional gain. The generated torque $T_{ff}$ and steering $\delta_f^*$, will be fed to the Command Interpreted Modeule (CIM) block to develop the desired C.G forces. The details of the torque-vectoring controller and functionality have been presented in \cite{Fallah1997}. In this formulation the optimal steering and velocity from the MPC are considered as two inputs for the torque-vectoring controller. The output of the controller is the driving/braking torque correction required for the stabilization of the vehicle. The desired C.G. forces derived from the CIM block are defined as:   
     \begin{equation}
       F_{des}=[F_x^*, F_y^*, G_z^*]^T
     \end{equation}  
     where $F_x^*$,$F_y^*$ and $G_z^*$ are the desired C.G. longitudinal and lateral forces and yaw moment, respectively. The actual C.G. forces of the vehicle are:
         \begin{equation}
     F=[F_x, F_y, G_z]^T
     \end{equation}  
     where $F_x$, $F_y$ and $G_z$ are the actual C.G. longitudinal and lateral forces and yaw moment of the vehicle, respectively. Each C.G. force components is a function of  longitudinal and lateral tire forces (see \cref{Force_conv}) as: 
            \begin{align}
           F_x=F_x(f_{x1},\dots, f_{x4}, f_{y1},\dots,f_{y4})\\
            F_y=F_y(f_{x1},\dots, f_{x4}, f_{y1},\dots,f_{y4})\\
            G_z=G_z(f_{x1},\dots, f_{x4}, f_{y1},\dots,f_{y4}) 
            \end{align} 
where, $f_{xi}$ and $f_{yi}$,$(i=1,\dots,4)$  are longitudinal and lateral tire forces on each wheel of the vehicle. The corresponding adjusted C.G. forces to minimize the error between the actual $F$ and desired, $F_{des}$, is represented by:
         \begin{equation}
           F(f+\theta f)\approx F(f)+\nabla F(f)\delta f
           \end{equation}   
where, $\nabla F(f)$ is the Jacobian matrix and $f$ is the total longitudinal and lateral forces on each wheel. The control action needed to minimize the error is:
  \begin{equation}
     \theta f=[\theta f_{x1},\dots,\theta f_{x4}, \theta f_{y1}, \dots, \theta f_{y4}]^T
 \end{equation}   
and is formulated as:  
 
   \begin{equation}
    \begin{gathered}
 \theta f=[W_{df}+W_{f}+(\nabla F(f)^TW_E)\nabla F(f)]^{-1}\\
 .[\nabla F(f)^T(W_EE)-W_{f}f]
  \end{gathered}  
 \end{equation}  

Where $W_{df}$, $W_f$ are weighing matrices. The applied corrective torque on each wheel is $\delta T=R_{eff}\times \theta f$. For brevity of the paper, further details about calculating control actions and weighting metrics are available in \cite{Fallah1997,holistic}.     
\section{Numerical Results}
\label{Results}
In this section, the closed-loop framework presented in \cref{controlsystem} is studied in a high-fidelity environment. The scenarios used to perform this study is as: $1)$ an obstacle avoidance when the subject vehicle is travelling at velocities in the range [60,80] $\SI{}{\kilo\metre\per\hour}$ with road radius of $\SI{750}{\metre}$, and $2)$ the application of torque-vectoring controller to evaluate the performance of the collision avoidance manoeuvre under different radius of the road with low surface friction. The purpose of the two tests is to illustrate the effectiveness of the proposed collision avoidance system on curve roads subject to environment constraints and disturbances using three MPC strategies discussed in \cref{Controllers}. The vehicle parameters are tabulated in \cref{parameters}. The simulation results are conducted using combine IPG carmaker and \MATLAB/Simulink software.
\begin{table}[b!] 
	\caption{Design parameters}
	\label{parameters}
	\begin{center}
		\begin{tabular}{cc}
			\hline
			\textbf{Symbols} & \textbf{Value(Unit)}\\
            \hline
			$l_f$ &$\SI{1.43}{\metre}$\\ 
	
		    $l_r$ &$\SI{1.21}{\metre}$\\
		
		    $M$ (Subject Vehicle mass) &$\SI{1360}{\kilogram}$\\
		
		    $I_z$ (Yaw moment of inertia) &$\SI{2050}{\kilogram\metre^2}$\\
	
		    $w_{lane}$ (Lane width)  &$\SI{5}{\metre}$\\
		    
		    $l_f$ (lead vehicle) &$\SI{1.5}{\metre}$\\
		    
		    $l_r$ (lead vehicle) &$\SI{1.7}{\metre}$\\
		    
		    $t_s$ (sampling time) &$\SI{0.1}{\s}$\\
		    \hline
		\end{tabular}
	\end{center}
\end{table}
\subsection{Collision Avoidance Under Different Velocities}
\begin{centering}
 	\begin{figure}[t!]
 	  \hspace*{-0.2in}
 		\centering
 		 		\includegraphics[width=1\columnwidth]{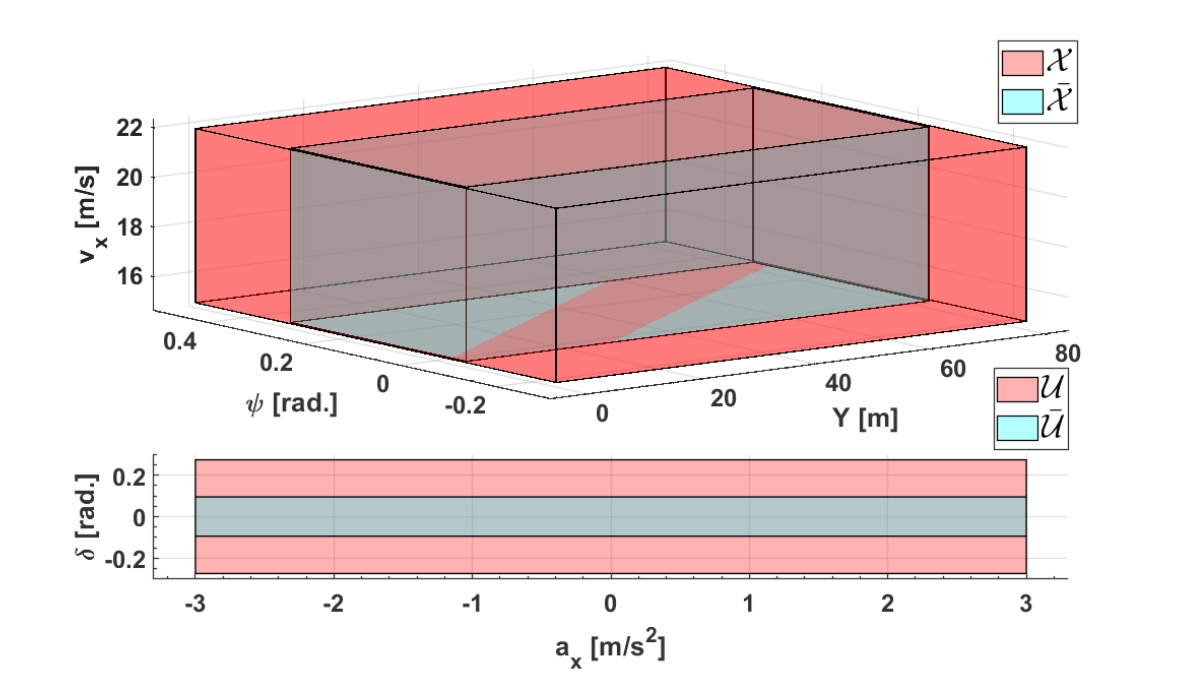}
 		\caption {State and input polyhedron and resulting tighten state and input}   
 		\label{poly}
 	\end{figure} 
 \end{centering}

\begin{centering}
 	\begin{figure}[b!]
 	  \hspace*{-0.6in}
 		\centering
 		\includegraphics[width=1.29\linewidth]{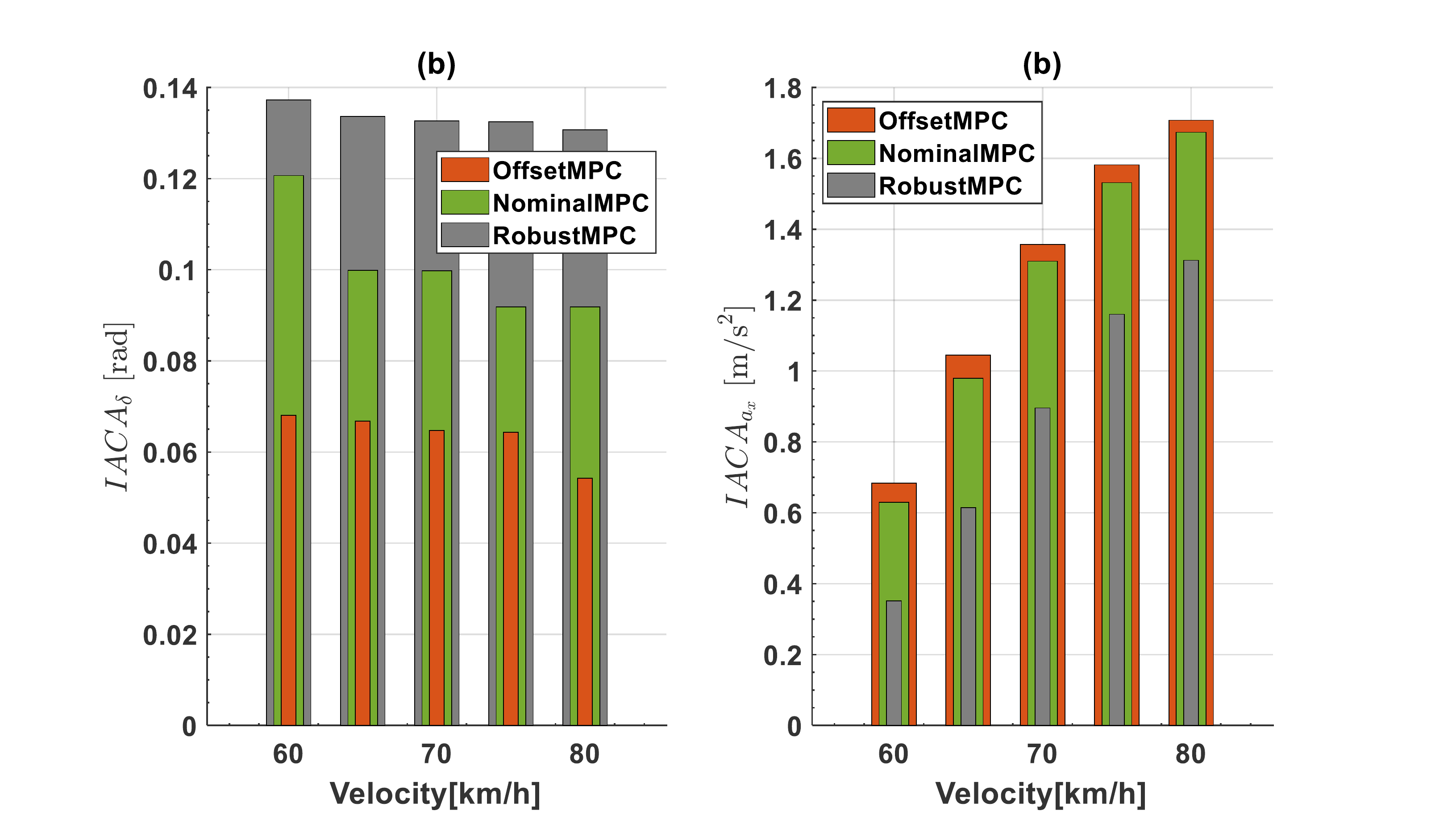}
 		\caption {Performance indicators of :{$(a)$ Steering, and $(b)$ Longitudinal acceleration}}    
 		\label{IACA-g}
 	\end{figure} 
 	\end{centering}

\begin{centering}
 	\begin{figure}[b!]
 	  \hspace*{-0.6in}
 		\centering
 		\includegraphics[width=1.29\linewidth]{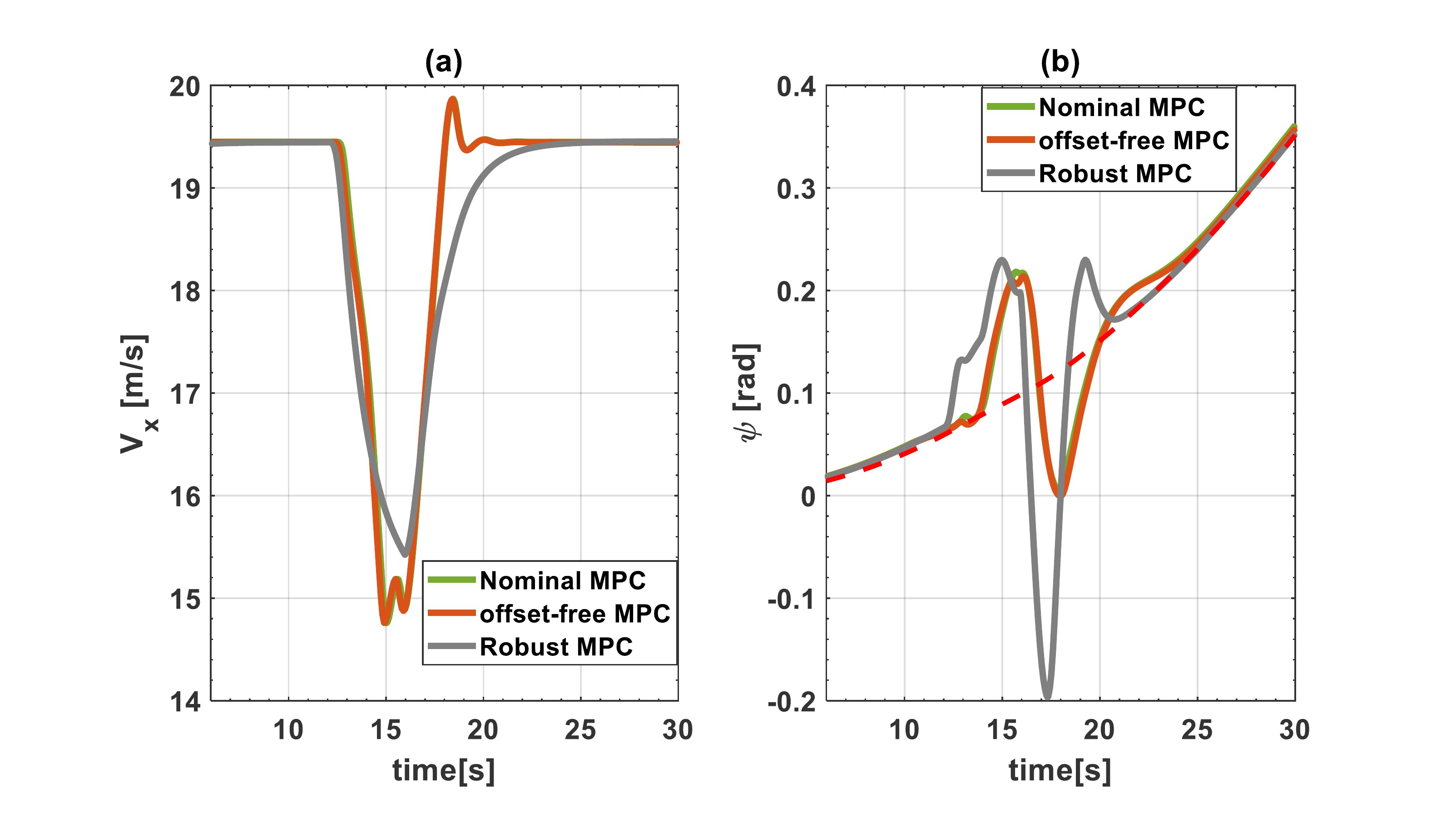}
 		\caption {$(a)$ Longitudinal velocity, and $(b)$ Heading angle \textbf{Note}: red dash line represents reference heading angle}    
 		\label{speedheading}
 	\end{figure} 
 	\end{centering}

The simulation environment is initialized with the subject vehicle behind the lead vehicle. Then the system plans the trajectory at each sampling time while applying FCC and RCC constraints based on the position of subject vehicle and the obstacle. The simulation results conducted for different velocities to assess the ability of the vehicle to perform the collision avoidance manoeuvre. Moreover, all predefined MPCs are initialized with short-sighted prediction horizon ($N=14$), for imitating more extreme and critical conditions in a collision avoidance scenario. The projection of the state and input constraints of kinematic vehicle model used in all MPCs are represented in \cref{poly} where $\mathcal{X}$ represents the state constraints of the nominal MPC and offset-free MPC while $\bar{\mathcal{X}}$ is tightened state which is used for robust MPC as state constraints.
\vspace*{-5mm} 
 \begin{center}
 	\begin{figure*}[t!]
   	\begin{multicols}{2}
   	\includegraphics[width=2\linewidth]{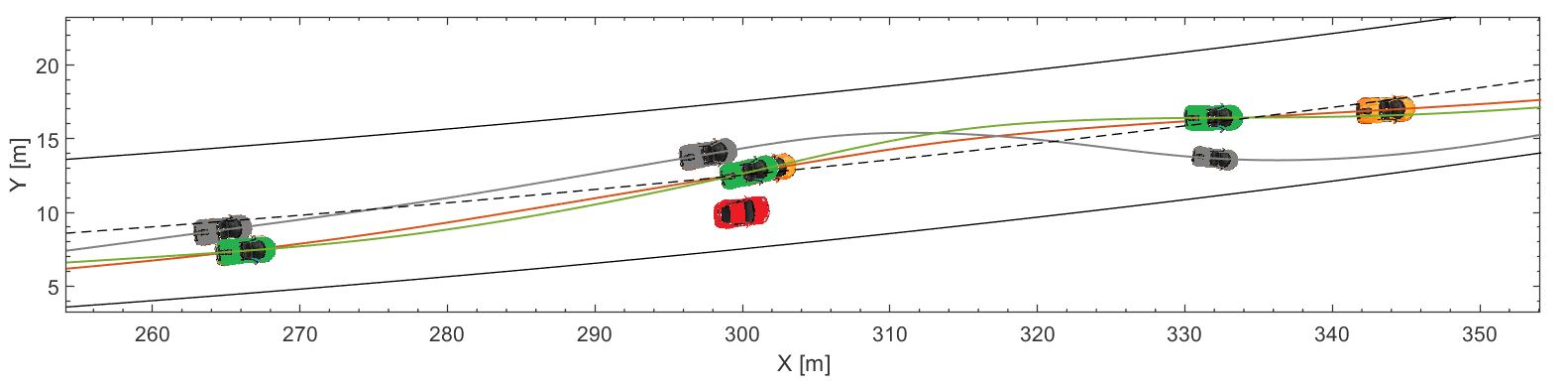}
   	\end{multicols}
   	\vspace*{-5mm}
   	\caption{Simulation results for collision avoidance test, \textbf{Note}: The orange line represents offset free MPC, Green line normal MPC and grey line robust MPC}
   	   \label{trajectory}
   \end{figure*}
\end{center} 
\vspace*{-3 mm} 
 Subsequently $\mathcal{U}$ is the input constraint for nominal MPC and offset-free MPC whereas $\bar{\mathcal{U}}$ is the result of tightened input set. For offset-free MPC the extra state constraints $ey$ and $e_{\psi}$ are chosen as $e_y$ = [-4,4]$\SI{}{\metre}$ and $e_{\psi}$ = [-0.15, 0.15]$rad$. It is noteworthy that the structure of robust positively invariant set ($\mathcal{Z}$) introduced in section \ref{RMPC}, has a dependence on size of disturbance set $\mathcal{W}$ and the matrix $A_K$. Since the only degree of freedom available to design $\mathcal{Z}$ is by means of matrix $A_K$, careful choose of the appropriate controller $K$ is necessary to ensure stable error dynamics, leading to calculate the tightened input and state constraints. An interested reader can refer to \cite{Dixit2019} for detailed explanation of the procedure. \cref{trajectory}, illustrates the obstacle avoidance scenario with initial speed of $\SI{80}{\kilo\metre\per\hour}$ with assumption of high friction surface road condition. As can be seen from this figure, all MPC strategies successfully avoid the collision from the lead vehicle (red car) and perform a complete collision avoidance manoeuvre. It is noteworthy, despite the fact that the nominal MPC and offset-free MPC can be effective for generating trajectory in collision avoidance scenario with a fixed longitudinal velocity, their performance can be degraded when the longitudinal speed is varying due to braking and acceleration while performing the manoeuvre. This can be confirmed in \cref{trajectory} where offset free MPC and nominal MPC generate a trajectory very close to the lead vehicle during initial lane change. On the other hand, the robust MPC based trajectory maintain the safety margins to the lead vehicle during the initial lane change. It is noteworthy, the parameters in MPCs (i.e $Q$, $R$ and $P$) can be tunned to adjust the aggressiveness of collision avoidance manoeuvre. To show the aggressiveness of all controllers, we compare the control efforts for the time span of collision avoidance manoeuvre as shown  \cref{IACA-g}. The control efforts are compared using the integral of the absolute value ($IACA$) of the steering and acceleration. The equations related to the performance indicator ($IACA$) can be written as \cite{Chatzikomis}:
     \begin{equation}     
     \begin{aligned}
               IACA_{\delta}&=\frac{1}{t_{fin}-t_{in}}\int_{t_{in}}^{t_{fin}} \left|{\delta}\right| dt\\
                IACA_{a_x}&=\frac{1}{t_{fin}-t_{in}}\int_{t_{in}}^{t_{fin}} \left|{a_x}\right| dt
            \label{IACA}
            \end{aligned}
            \end{equation}    
It is worth mentioning that as the velocity increases, the control effort to perform the collision avoidance manoeuvre is less compared to lower speed for all controllers. It is well known that the steady-state yaw-rate response to steering angle of a vehicle increases as the vehicle's velocity increases. Hence, as the velocity rises, smaller steering angle inputs are required to generate the same yaw-rate (for evading the fixed size obstacle) and this can be observed in \cref{IACA-g}(a). The \cref{Tableerror} illustrates the trend, where $IACA_{\delta}$ is at the maximum when the vehicle speed is $\SI{60}{\kilo\metre\per\hour}$, and reaches to its minimum value in high speed for all controllers. The result of the steering effort is reflected on the heading angle of the vehicle (in \cref{speedheading}(b)) where large evolution of heading angle is shown for robust MPC during either of the lane changes. It is noted that \cref{speedheading} is the time series of velocity and heading angle of the vehicle for all controllers for a velocity value of $\SI{75}{\kilo\metre\per\hour}$ as an example. The second performance indicator in \eqref{IACA} represents control effort on longitudinal acceleration. \cref{IACA-g}(b) compares the acceleration needed to perform collision avoidance manoeuvre for all controllers. It can be seen that, the longitudinal acceleration for nominal MPC and offset-free MPC is almost the same, whereas the longitudinal acceleration effort for robust MPC is lower than the rest. This can be confirmed in \cref{Tableerror} where the longitudinal acceleration for robust MPC is almost 40\% less than the rest of the controllers for all velocity values. The significant of this analysis can be seen in \cref{speedheading}(a) where the longitudinal velocity profile for offset-free and nominal MPC show larger evolution with overshoot during braking and accelerating. On the other hand, robust MPC demonstrates smooth profile without any high-frequency oscillation or overshoot. This can be inferred, where the robust MPC tackles the variations of the longitudinal vehicle speed, resulting in a smooth and jerk free braking/accelerating while performing collision avoidance manoeuvre. This analysis has been carried out to demonstrates the benefits of utilizing each controllers based on the users need, and the trade-off between control actions required for smoothness and handling of the vehicle in collision avoidance manoeuvre.

 It is beneficial to observe the maximum lateral error between the generated trajectory and the reference trajectory that vehicle needs to track. \cref{ey} represents maximum lateral errors for all controllers during and after performing obstacle avoidance. \cref{ey}(a) indicates that the lateral errors are gradually increasing as velocity of the vehicle increases. It can be seen that the maximum lateral error for nominal MPC, offset-free MPC and robust MPC reaches to the maximum of $\SI{0.34}{\metre}$, $\SI{0.18}{\metre}$ and $\SI{0.28}{\metre}$ when the speed is at $\SI{80}{\kilo\metre\per\hour}$. It also shows the superior performance of offset-free MPC in path tracking compared to other controllers. This  suggests that the accurate tracking requires the incorporation of path curvature into the system dynamics. \cref{ey}b shows the distance of the subject vehicle from the reference line during the obstacle avoidance manoeuvre for different velocities to demonstrates the safe distance from outer road boundary. In this figure, the capability of all MPC's in maintaining safe distance from the outer road boundary while performing collision avoidance is shown. 

\begin{centering}
 	\begin{figure}[t!]
 	  \hspace*{-0.6in} 
 		\centering
 		\includegraphics[width=4.5in,height=3in]{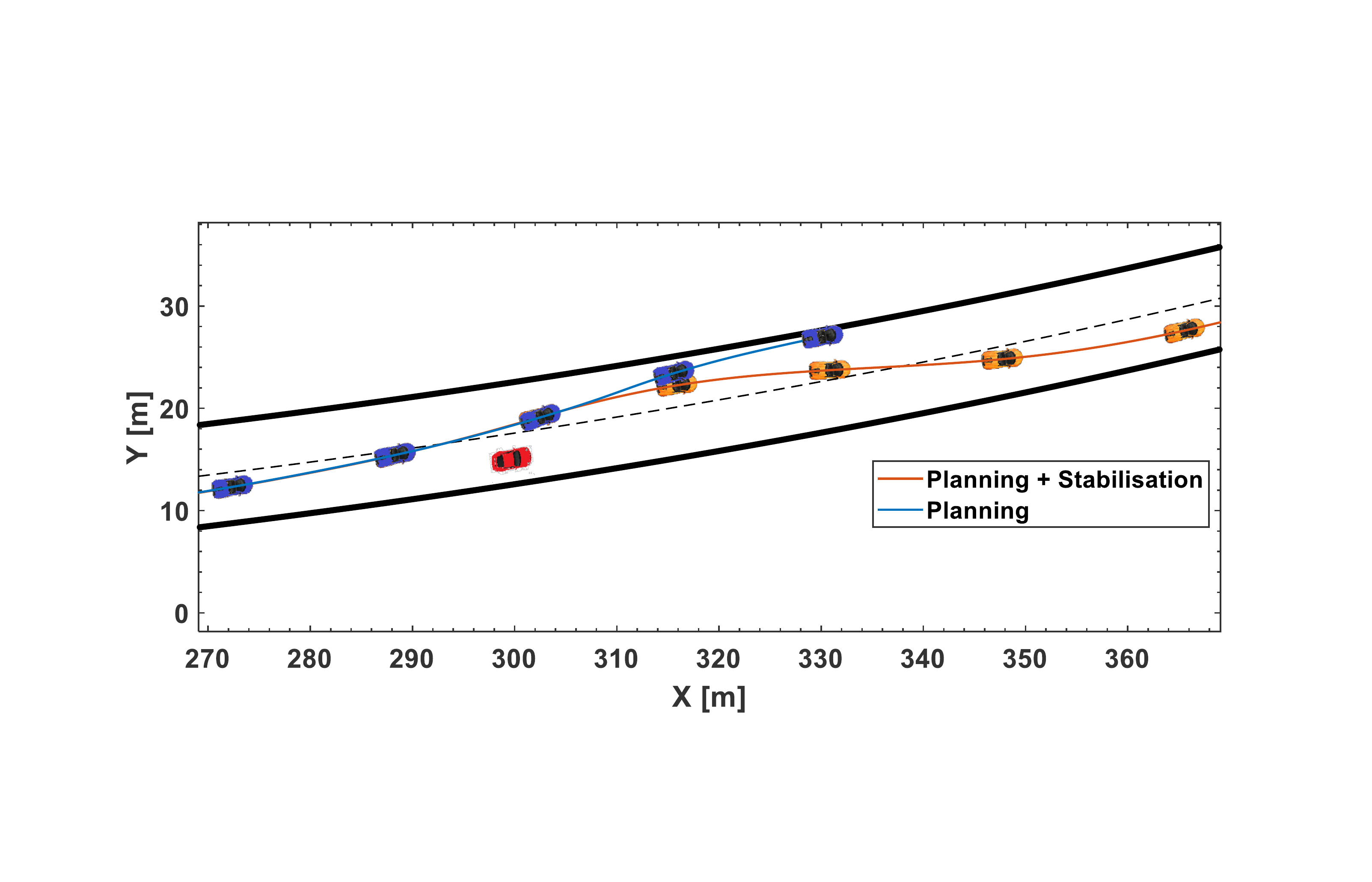}
 		\vspace*{-15mm}
 		\caption {Collision avoidance test under low friction surface and tight curvature. The orange line represents subject vehicle with torque-vectoring controller, and blue line without torque-vectoring controller}   
 		\label{CA_TV}
 	\end{figure} 
 	\end{centering}

\begin{centering}
 	\begin{figure}[b!]
 	  \hspace*{-0.44in} 
 		\centering
 		\includegraphics[width=1.2\linewidth]{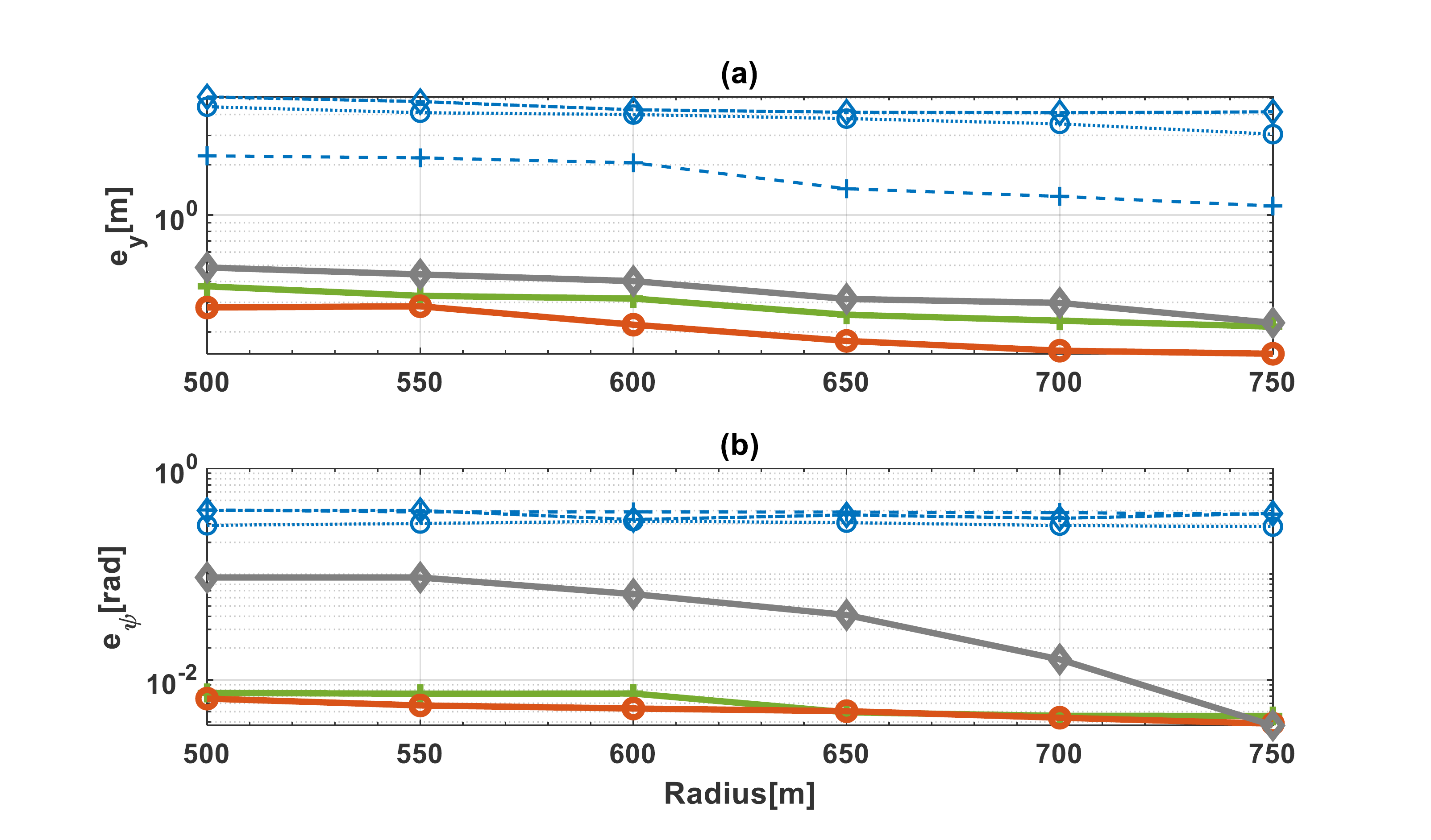}
 		\caption {Collision avoidance test under low friction surface for different value of curvature. $e_y$ represents lateral error and $e_\psi$ lateral heading angle error, \textbf{Note}: Dash blue lines represent the vehicle without torque-vectoring}   
 		\label{CA_TV_error}
 	\end{figure} 
 	\end{centering}

\begin{centering}
 	\begin{figure}[t!]
 	  \hspace*{-0.39in} 
 		\centering
 		\includegraphics[width=1\linewidth]{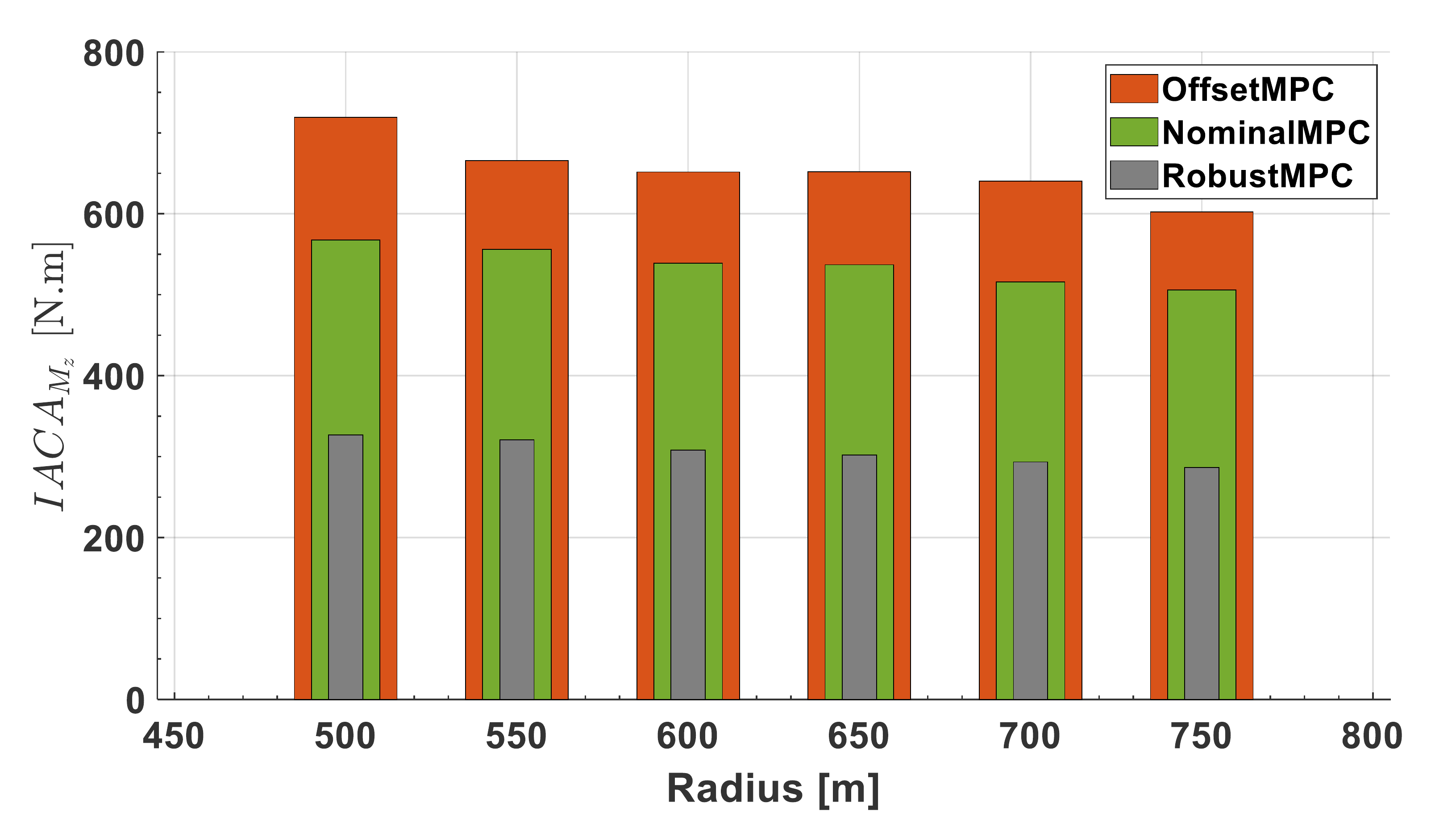}
 		\caption {Net control action from torque-vectoring controller}   
 		\label{CA_TV_overal}
 	\end{figure} 
 	\end{centering}

\begin{centering}
 	\begin{figure}[t!]
 	  \hspace*{-0.55in} 
 		\centering
 		\includegraphics[width=1.28\linewidth]{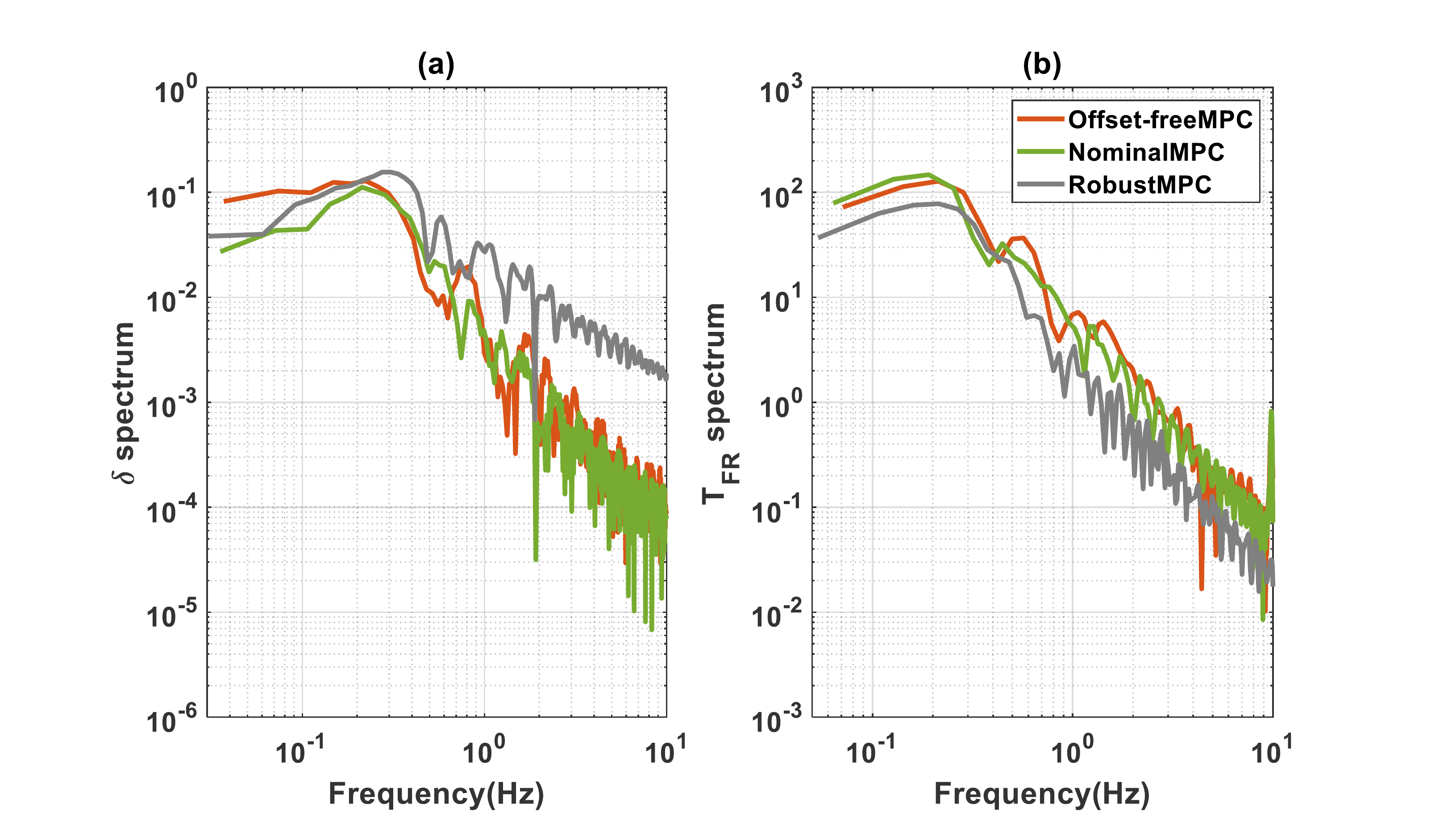}
 		\caption {Spectrum of (a) steering (b) front-right torque}   
 		\label{CA_TV_spec}
 	\end{figure} 
 	\end{centering}

 \footnotesize 
\begin{table}[b!]
	\caption{\small Performance indicators}
		\begin{tabular}{c||m{1cm}|m{1cm}|m{1cm}|m{1cm}|m{0.5cm}}
		\hline\hline
	\label{Tableerror}
	\multirow{2}{*} \textbf{NomMPC} & \textbf{$60km/h$}& \textbf{$65km/h$}& \textbf{$70km/h$}& \textbf{$75km/h$}& \textbf{$80km/h$}\\
            \hline
			$IACA_{\delta}(rad)$ &$0.1207$& $0.09987$&$0.09977$&$0.09183$&$0.09183$\\ 
	        \hline
	        $IACA_{a_x}(\frac{m}{s^2})$&0.6296&0.97&1.31&1.5&1.67\\
	        \hline

		    \hline\hline
		    \multirow{2}{*} \textbf{OffMPC} & && & & \\
            \hline
			$IACA_{\delta}(rad)$ &0.068& 0.066&0.0647&0.0643&0.05\\ 
	        \hline
	        $IACA_{a_x}(\frac{m}{s^2})$&0.68&1.04&1.35&1.58&1.7\\
	        \hline
	      
	          \hline\hline
	        		    \multirow{2}{*} \textbf{RobMPC} & && & & \\
            \hline
			$IACA_{\delta}(rad)$ &0.1373& 0.1337&0.1326&0.1324&0.1307\\ 
	        \hline
	        $IACA_{a_x}(\frac{m}{s^2})$&0.35&0.61&0.89&1.16&1.31\\
	        \hline
	     
	        \hline\hline
		    	 	 \end{tabular}
            \end{table}
          \normalsize  
   \begin{centering}
 	
\subsection{Collision Avoidance using Torque Vectoring}
Torque-vectoring controller enhances vehicle's handling and safety in extreme manoeuvres such as low surface friction or high road curvature. To show the need for torque-vectoring controller, this section evaluates the performance of all MPCs when they are combined with torque vectoring controller to maintain the stability of the vehicle under different curvature as well as environment conditions. For evaluation purposes, the performance of trajectory planning controllers is compared with the case of vehicle without having torque-vectoring controller. 
 	\begin{figure}[t!]
	\hspace*{-0.6in}
	\centering
	\includegraphics[width=1.28\linewidth]{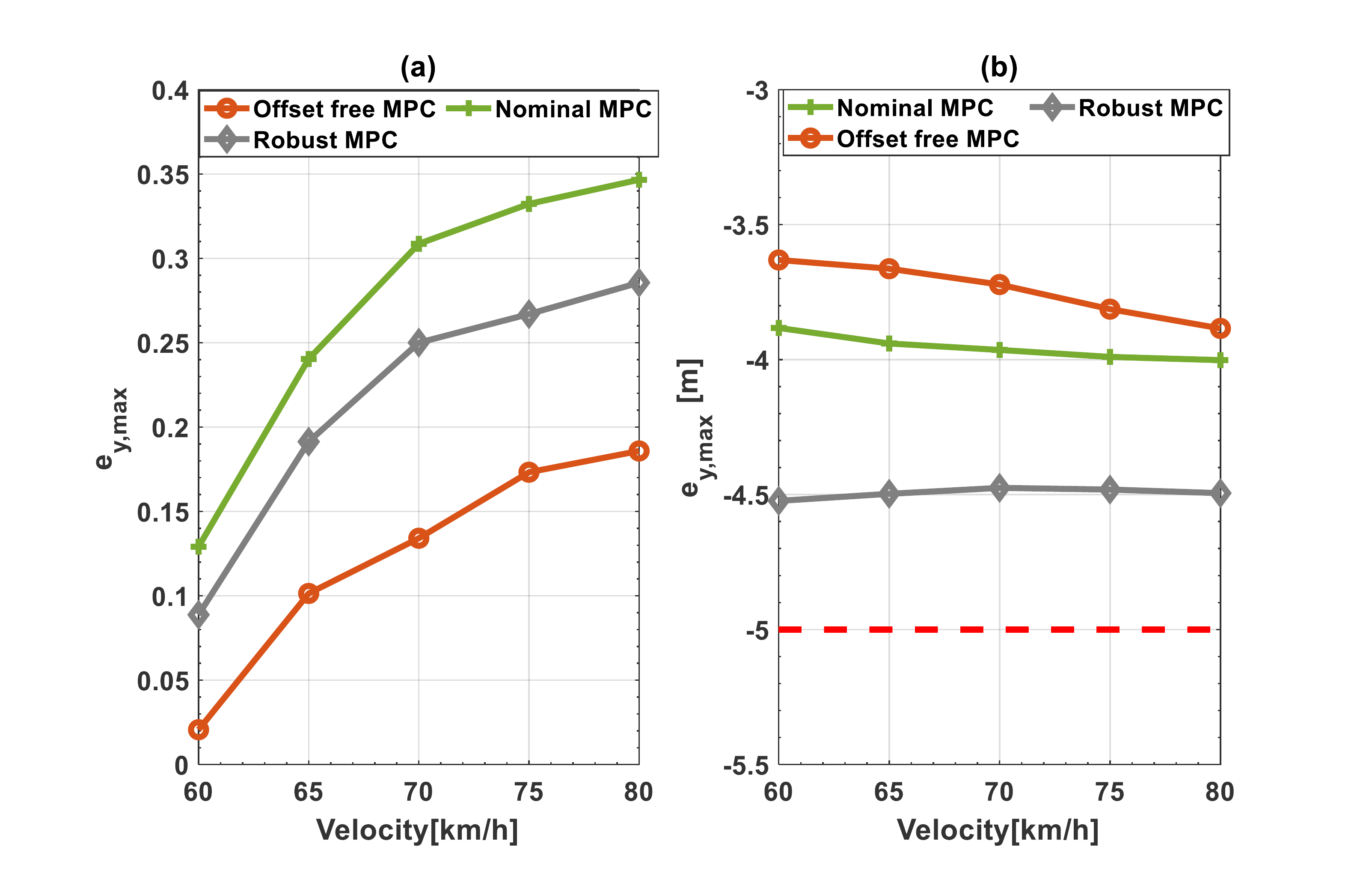}
	\caption {Maximum lateral error $(a)$Tracking error, and $(b)$  Maximum lateral error during obstacle avoidance, \textbf{Note}: Red dash line indicates outer road edge}   
	\label{ey}
\end{figure} 
\end{centering}
As an example, \cref{CA_TV}, shows the performance of the offset-free MPC with and without torque-vectoring controller. In this example, simulation result has been carried out under tight curvature with radius, $R=\SI{500}{\metre}$ and friction surface of $\mu=0.4$. This scenario represents a harsh driving condition for performing collision avoidance manoeuvre. As can be seen, controller with torque-vectoring, stabilizes the vehicle during collision avoidance manoeuvre with initial speed of $v_x=\SI{75}{\kilo\metre\per\hour}$. The figure also shows the incapability of the controller without torque-vectoring in stabilizing the vehicle in such harsh driving manoeuvre.\\  A parametric analysis carried out to verify the efficacy of the proposed combined framework after performing collision avoidance manoeuvre to demonstrate the tracking capability of the controllers under low surface friction for different radius of the path. In this analysis, The maximum lateral and heading error (\cref{CA_TV_error}) between the reference and actual trajectory of the vehicle is investigated. The maximum lateral error for all controllers increases gradually as the radius of the road decreases (curvature increase). Among all controllers, offset-free MPC experiences a better reference tracking. This can be confirmed in \cref{Tableerror-sec} where the largest lateral error can be seen with radius of $R=\SI{500}{\metre}$ with $e_y=\SI{0.27}{\metre}$, following with nominal and robust MPC with $e_y=\SI{0.37}{\metre}$ and $e_y=\SI{0.48}{\metre}$ respectively. \cref{CA_TV_error} indicates that, MPCs incorporating torque-vectoring controller maintain vehicle from further deviation of reference trajectory and keep the vehicle within the lane limits. However, vehicle without torque-vectoring controller (blue dashes in \cref{CA_TV_error}a) starts to deviate from reference path with above $\SI{1}{\metre}$ deviation.      
Similarly, for the heading angle error ($e_\psi$) in \cref{CA_TV_error}b, all controllers maintain the directionality of the vehicle by keeping the heading angle error to be small, allowing subject vehicle to follow the reference heading angle. On the other hand, controllers without torque-vectoring control experience large lateral and heading error resulting in loss of stability and directionality (blue dash lines in \cref{CA_TV_error}). It is noted that when the vehicle travels on a moderate driving condition, i.e., low speed or low-curvature, the path tracking objectives are fulfilled. However, as the vehicle travels in an extreme conditions i.e., high-curvature or high speed, the required heading angle may deteriorate the tracking capability. The reason of this phenomena is twofold: (1) there would be unavoidable tire sliding effects, (2) presence of sideslip angle of the vehicle  which is not included in the kinematic vehicle model. Therefore, as the radius of the road decreases (curvature increases), the maximum heading angle error will increase. It is noted that, although the largest heading error happens when the radius of the road is $\SI{500}{\metre}$ where curvature is tight, the MPC's with incorporation of torque-vectoring controller maintain the directionality of the vehicle. This can be seen in \cref{Tableerror-sec} where the largest heading error for nominal MPC, offset-free MPC and robust MPC is 0.0075rad, 0.0066rad and 0.09rad respectively.\\
The net control action from all tire forces can be seen in \cref{CA_TV_overal}. In this figure, The control effort ($IACA_{M_z}$) needed for stabilizing the vehicle while perform the collision avoidance manouvre, for offset MPC and nominal MPC is larger repsect to robust MPC. Robust MPC incorporates the worst-case disturbance realizations while solving the optimisation problem. In the robust MPC optimization problem, the performance index is being optimized allows MPC to operate between lower and upper bound of disturbance set. As a consequence, control law maintain the system within an invariant tube around nominal MPC and attempt to execute the control action more aggressively to steer the uncertain system towards the reference system. The results of this steering control action, allows the intervention of the torque-vectoring controller, for applying additional torque, to be limited. On the other hand offset-free MPC and nominal MPC require extra torque actuation for stabilisation. This can be confirmed in \cref{CA_TV_spec} where less steering action is required in both low frequency and high frequency range compared to robust MPC. 
The tabulated results from the performance indicator for the torque-vectoring actuation as well as maximum lateral and heading error values are reported in \cref{Tableerror-sec}. It is shown that, in driving scenario where radius is 750m, the combined controllers can enhanced vehicle stability while performing collision avoidance with minimum lateral and heading errors. However, as radius of the road decreases, extra additional yaw moment from torque-vectoring controller is needed in order to stabilize the vehicle. This table summarized the performance of the combined controllers in terms of tracking the reference values from the road. It can be inferred from the table the need of torque vectoring controller for stabilization under extreme driving condition and safe driving condition on different curvature.
 \footnotesize
\begin{table}[b!]
	\caption{\small Performance indicator and maximum lateral and heading error}
		\begin{tabular}{c||m{0.7cm}|m{0.7cm}|m{0.7cm}|m{0.7cm}|m{0.7cm}|m{0.7cm}}
		\hline\hline
	\label{Tableerror-sec}
	\multirow{2}{*} \textbf{NomMPC} & \textbf{$500m$}& \textbf{$550m$}& \textbf{$600m$}& \textbf{$650m$}& \textbf{$700m$}& \textbf{$750m$}\\
            \hline
			$IACA_{M_z}(N.m)$ &567.5& 556.1&539.1&537.1&515.9&506\\ 
	        \hline
	        $e_y(m)$&0.3745&0.328&0.316&0.253&0.233&0.214\\
	        \hline
	        $e_{\psi}(rad)$ &0.0075& 0.0073&0.0074&0.0049&0.00456&0.00454\\
		    \hline\hline
		    \multirow{2}{*} \textbf{OffMPC} & && & & \\
            \hline
			$IACA_{M_z}(N.m)$ &719.1&665.8&651.7&651.9&640.4&602.2\\ 
	        \hline
	        $e_y(m)$&0.27& 0.28&0.22&0.17&0.154&0.147\\
	        \hline
	        $e_{\psi}(rad)$ &0.0066&0.0057&0.0053&0.0050&0.0043&0.0038\\
	          \hline\hline
	        		    \multirow{2}{*} \textbf{RobMPC} & && & & \\
            \hline
			$IACA_{M_z}(N.m)$ &326.7&320.5&308.2&302&293.4&286.6\\ 
	        \hline
	        $e_y(m)$&0.48&0.44&0.40&0.31&0.29&0.22\\
	        \hline
	        $e_{\psi}(rad)$ &0.09& 0.089&0.064&0.041&0.015&0.0036\\
	        \hline\hline
		    	 	 \end{tabular}
            \end{table}
          \normalsize       

\section{Conclusion}
\label{Conclusion}
This paper, compared the performance of three different MPC's for trajectory planning/control, combined with torque vectoring controller for collision avoidance tests. A novel path planning was designed based on model predictive control algorithm. This formulation, deals with non-convex collision avoidance constraints on curve road, allowing the use of effective optimisation solvers for computing vehicle trajectory through MPC based strategies. In order to implement these constraints, a technique was presented to have feasible and convex free optimization on a curve road. This allows, the trajectory planning controller to generate safe and feasible trajectories with admissible inputs while performing collision avoidance manoeuvre. Furthermore, Three different MPCs, entitled as nominal MPC, robust MPC and offset-free MPC has been compared. It was seen from the simulation results that offset-free MPC outperform the rest of the controllers in terms of driving smoothness and path tracking. However, robust MPC add further benefit for designing collision avoidance system, by providing extra safety margin. Additionally, torque vectoring algorithm has been combined with all MPCs through its separate implementation. To assess the performance of the combined controller, different value of curvature as well as low surface friction have been used to simulate the controllers under extreme driving conditions. The numerical results in Simulink/IPG CarMaker co-simulation environment demonstrated that, the combined controllers improved vehicle performance under extreme and dynamic environment, compare to controllers without torque-vectoring, and fulfil the safety considerations for collision avoidance manoeuvre.


%

\FloatBarrier

\appendices




\ifCLASSOPTIONcaptionsoff
  \newpage
\fi



%
\bibliographystyle{IEEEtran}
\bibliography{Journal_2019}

\begin{thebibliography}{10}
\providecommand{\url}[1]{#1}
\csname url@samestyle\endcsname
\providecommand{\newblock}{\relax}
\providecommand{\bibinfo}[2]{#2}
\providecommand{\BIBentrySTDinterwordspacing}{\spaceskip=0pt\relax}
\providecommand{\BIBentryALTinterwordstretchfactor}{4}
\providecommand{\BIBentryALTinterwordspacing}{\spaceskip=\fontdimen2\font plus
\BIBentryALTinterwordstretchfactor\fontdimen3\font minus
  \fontdimen4\font\relax}
\providecommand{\BIBforeignlanguage}[2]{{%
\expandafter\ifx\csname l@#1\endcsname\relax
\typeout{** WARNING: IEEEtran.bst: No hyphenation pattern has been}%
\typeout{** loaded for the language `#1'. Using the pattern for}%
\typeout{** the default language instead.}%
\else
\language=\csname l@#1\endcsname
\fi
#2}}
\providecommand{\BIBdecl}{\relax}
\BIBdecl

\bibitem{Funke2017}
J.~Funke, M.~Brown, S.~M. Erlien, and J.~C. Gerdes, ``{Collision Avoidance and
  Stabilization for Autonomous Vehicles in Emergency Scenarios},'' \emph{IEEE
  Transactions on Control Systems Technology}, vol.~25, no.~4, pp. 1204--1216,
  2017.

\bibitem{Cheng2019}
S.~Cheng, L.~Li, H.-Q. Guo, Z.-G. Chen, and P.~Song, ``{Longitudinal Collision
  Avoidance and Lateral Stability Adaptive Control System Based on MPC of
  Autonomous Vehicles},'' \emph{IEEE Transactions on Intelligent Transportation
  Systems}, vol.~PP, pp. 1--10, 2019.

\bibitem{Xu2019}
S.~Xu and H.~Peng, ``{Design, Analysis, and Experiments of Preview Path
  Tracking Control for Autonomous Vehicles},'' \emph{IEEE Transactions on
  Intelligent Transportation Systems}, vol.~PP, pp. 1--11, 2019.

\bibitem{Boyali2018}
A.~Boyali, V.~John, Z.~Lyu, R.~Swarn, and S.~Mita, ``{Self-scheduling robust
  preview controllers for path tracking and autonomous vehicles},'' \emph{2017
  Asian Control Conference, ASCC 2017}, vol. 2018-Janua, pp. 1829--1834, 2018.

\bibitem{Wang2018a}
\BIBentryALTinterwordspacing
Z.~Wang, U.~Montanaro, S.~Fallah, A.~Sorniotti, and B.~Lenzo, ``{A gain
  scheduled robust linear quadratic regulator for vehicle direct yaw moment
  Control},'' \emph{Mechatronics}, vol.~51, no. January, pp. 31--45, 2018.
  [Online]. Available: \url{https://doi.org/10.1016/j.mechatronics.2018.01.013}
\BIBentrySTDinterwordspacing

\bibitem{Khosravani2018}
S.~Khosravani, M.~Jalali, A.~Khajepour, A.~Kasaiezadeh, S.~K. Chen, and
  B.~Litkouhi, ``{Application of lexicographic optimization method to
  integrated vehicle control systems},'' \emph{IEEE Transactions on Industrial
  Electronics}, vol.~65, no.~12, pp. 9677--9686, 2018.

\bibitem{Stentz1995}
A.~Stentz, ``{The Focussed D* Algorithm for Real-Time Replanning},'' in
  \emph{14th international joint conference on artificial intelligence}, San
  Francisco, California, 1995, pp. 1652--1659.

\bibitem{Viking2007}
M.~L. {Dave Ferguson, Thomas M.howard}, ``{Motion Planning in
  UrbanEnvironments},'' \emph{Journal of Field Robotics}, vol.~25, pp.
  939--960., 2008.

\bibitem{Chen2019}
\BIBentryALTinterwordspacing
Y.~Chen, H.~Ye, and M.~Liu, ``{Hierarchical Trajectory Planning for Autonomous
  Driving in Low-speed Driving Scenarios Based on RRT and Optimization},''
  2019. [Online]. Available: \url{http://arxiv.org/abs/1904.02606}
\BIBentrySTDinterwordspacing

\bibitem{Franze2016}
G.~Franze and W.~Lucia, ``{A Receding Horizon Control Strategy for Autonomous
  Vehicles in Dynamic Environments},'' \emph{IEEE Transactions on Control
  Systems Technology}, vol.~24, no.~2, pp. 695--702, 2016.

\bibitem{p8}
Y.~Gao, A.~Gray, J.~V. Frasch, T.~Lin, E.~Tseng, J.~K. Hedrick, and
  F.~Borrelli, ``{Spatial Predictive Control for Agile Semi-Autonomous Ground
  Vehicles},'' \emph{Proceedings of the 11th International Symposium on
  Advanced Vehicle Control}, vol. VD11, no.~2, pp. 1--6, 2012.

\bibitem{Karlsson2016}
J.~Karlsson, N.~Murgovski, and J.~Sj{\"{o}}berg, ``{Temporal vs. Spatial
  formulation of autonomous overtaking algorithms},'' \emph{IEEE Conference on
  Intelligent Transportation Systems, Proceedings, ITSC}, pp. 1029--1034, 2016.

\bibitem{Nilsson2015a}
\BIBentryALTinterwordspacing
J.~Nilsson, P.~Falcone, M.~Ali, and J.~Sj{\"{o}}berg, ``{Receding horizon
  maneuver generation for automated highway driving},'' \emph{Control
  Engineering Practice}, vol.~41, pp. 124--133, 2015. [Online]. Available:
  \url{http://dx.doi.org/10.1016/j.conengprac.2015.04.006}
\BIBentrySTDinterwordspacing

\bibitem{DeNovellis2015}
\BIBentryALTinterwordspacing
L.~{De Novellis}, A.~Sorniotti, P.~Gruber, J.~Orus, J.~M. {Rodriguez Fortun},
  J.~Theunissen, and J.~{De Smet}, ``{Direct yaw moment control actuated
  through electric drivetrains and friction brakes: Theoretical design and
  experimental assessment},'' \emph{Mechatronics}, vol.~26, pp. 1--15, 2015.
  [Online]. Available:
  \url{http://dx.doi.org/10.1016/j.mechatronics.2014.12.003}
\BIBentrySTDinterwordspacing

\bibitem{Lu2016}
\BIBentryALTinterwordspacing
Q.~Lu, P.~Gentile, A.~Tota, A.~Sorniotti, P.~Gruber, F.~Costamagna, and J.~{De
  Smet}, ``{Enhancing vehicle cornering limit through sideslip and yaw rate
  control},'' \emph{Mechanical Systems and Signal Processing}, vol.~75, pp.
  455--472, 2016. [Online]. Available:
  \url{http://dx.doi.org/10.1016/j.ymssp.2015.11.028}
\BIBentrySTDinterwordspacing

\bibitem{Taherian2019}
S.~Taherian, U.~Montanaro, S.~Dixit, and S.~Fallah, ``{Integrated Trajectory
  Planning and Torque Vectoring for Autonomous Emergency Collision
  Avoidance},'' \emph{International Conference on Intelligent Transportation
  Systems (ITSC)}, 2019.

\bibitem{Kong}
J.~Kong, M.~Pfeiffer, G.~Schildbach, and F.~Borrelli, ``{Kinematic and Dynamic
  Vehicle Models for Autonomous Driving Control Design},'' in \emph{Kinematic
  and Dynamic Vehicle Models for Autonomous Driving Control Design}, 2015.

\bibitem{Borrelli2007}
F.~Borrelli and M.~Morari, ``{Offset free model predictive control},''
  \emph{Proceedings of the IEEE Conference on Decision and Control}, pp.
  1245--1250, 2007.

\bibitem{Chatzikomis}
C.~Chatzikomis, A.~Sorniotti, P.~Gruber, M.~Zanchetta, and D.~Willans,
  ``{Comparison of Path Tracking and Torque- Vectoring Controllers for
  Autonomous Electric Vehicles},'' \emph{IEEE}, vol.~3, no.~4, pp. 559--570,
  2018.

\bibitem{p12}
\BIBentryALTinterwordspacing
J.~Nilsson, Y.~Gao, A.~Carvalho, and F.~Borrelli, ``{Manoeuvre generation and
  control for automated highway driving},'' \emph{IFAC Proceedings Volumes},
  vol.~47, no.~3, pp. 6301--6306, 2014. [Online]. Available:
  \url{http://linkinghub.elsevier.com/retrieve/pii/S1474667016426005}
\BIBentrySTDinterwordspacing

\bibitem{Gao2014a}
Y.~Gao, A.~Gray, A.~Carvalho, H.~E. Tseng, and F.~Borrelli, ``{Robust nonlinear
  predictive control for semiautonomous ground vehicles},'' \emph{Proceedings
  of the American Control Conference}, pp. 4913--4918, 2014.

\bibitem{Plessen2018}
M.~G. Plessen, P.~F. Lima, J.~Martensson, A.~Bemporad, and B.~Wahlberg,
  ``{Trajectory planning under vehicle dimension constraints using sequential
  linear programming},'' \emph{IEEE Conference on Intelligent Transportation
  Systems, Proceedings, ITSC}, vol. 2018-March, pp. 1--6, 2018.

\bibitem{Scheuer1997}
A.~Scheuer and T.~Fraichard, ``{Collision-free and continuous-curvature path
  planning for car-like robots},'' \emph{Proceedings - IEEE International
  Conference on Robotics and Automation}, vol.~1, no. April, pp. 867--873,
  1997.

\bibitem{Shin1992}
D.~Shin and S.~Singh, ``{Path Generation for a Robot Vehicle Using Composite
  Clothoid Segments},'' Tech. Rep., 1990.

\bibitem{Mayne2010}
J.~B.~R. Mayne and D.~Q., ``{Model Predictive Control: Theory and Design},'' in
  \emph{Model Predictive Control: Theory and Design}, 2015, pp. 237--241.

\bibitem{Dixit2019}
S.~Dixit, U.~Montanaro, M.~Dianati, D.~Oxtoby, T.~Mizutani, A.~Mouzakitis, and
  S.~Fallah, ``{Trajectory Planning for Autonomous High-Speed Overtaking in
  Structured Environments Using Robust MPC},'' \emph{IEEE Transactions on
  Intelligent Transportation Systems}, pp. 1--14, 2019.

\bibitem{Rakovic2005}
S.~V. Rakovic, E.~C. Kerrigan, K.~I. Kouramas, and D.~Q. Mayne, ``{Invariant
  Approximations of the Minimal Robust Positively Invariant Set},'' \emph{IEEE
  TRANSACTIONS ON AUTOMATIC CONTROL}, vol.~50, no.~3, pp. 406--410, 2005.

\bibitem{Gao2014}
\BIBentryALTinterwordspacing
Y.~Gao, A.~Gray, H.~E. Tseng, and F.~Borrelli, ``{A tube-based robust nonlinear
  predictive control approach to semiautonomous ground vehicles},''
  \emph{Vehicle System Dynamics}, vol.~52, no.~6, pp. 802--823, 2014. [Online].
  Available: \url{https://doi.org/10.1080/00423114.2014.902537}
\BIBentrySTDinterwordspacing

\bibitem{Jalali2012}
K.~Jalali, S.~Lambert, and J.~McPhee, ``{Development of a Path-following and a
  Speed Control Driver Model for an Electric Vehicle},'' \emph{SAE
  International Journal of Passenger Cars - Electronic and Electrical Systems},
  vol.~5, no.~1, pp. 100--113, 2012.

\bibitem{Fallah1997}
S.~Fallah and A.~Khajepour, ``{Vehicle Optimal Torque Vectoring Using
  State-Derivative Feedback and Linear Matrix Inequality},'' \emph{IEEE
  TRANSACTIONS ON VEHICULAR TECHNOLOGY}, vol.~62, no.~4, 2013.

\bibitem{holistic}
S.-k. Chen, S.-k. Chen, Y.~Ghoneim, N.~Moshchuk, and B.~Litkouhi, ``{Tire
  Force-based Holistic Corner Control Tire Force-based Holistic Corner
  Control},'' no. November, 2012.

\end{thebibliography}




%

\begin{IEEEbiography}[{\includegraphics[width=1in,height=2in,clip,keepaspectratio]{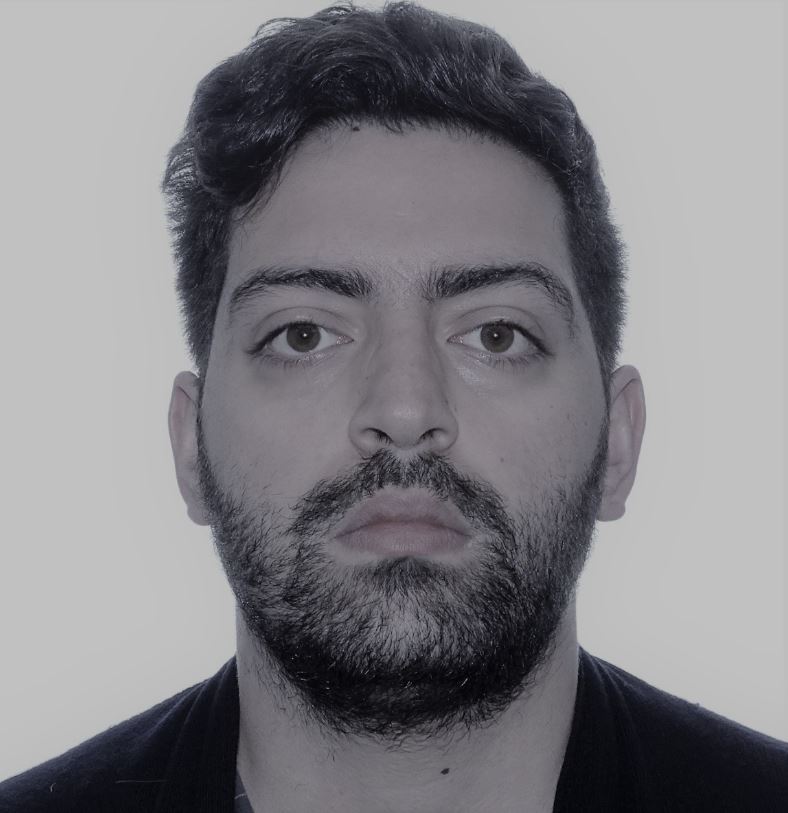}}]
{Shayan Taherian} received the M.Sc. degree in automation and control engineering from the Polytechnic of Milan University, Italy, in 2016. He is currently pursuing the Ph.D. degree with the
Centre of Automotive Engineering, University of
Surrey, Guildford, U.K. His research interests are
vehicle dynamics and control, trajectory planning for autonomous vehicles, and intelligent
vehicles.
\end{IEEEbiography}
\begin{IEEEbiography}[{\includegraphics[width=1in,height=2in,clip,keepaspectratio]{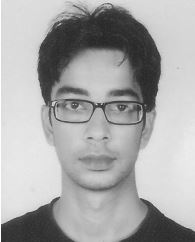}}]{Shilp Dixit} received the M.Sc. degree in automotive technology from the Eindhoven University of
Technology, Eindhoven, The Netherlands, in 2015.
He currently finished the Ph.D. degree with the
Centre of Automotive Engineering, University of
Surrey, Guildford, U.K. His research interests are
vehicle dynamics and control, trajectory planning
and tracking for autonomous vehicles, and intelligent
vehicles.
\end{IEEEbiography}
\begin{IEEEbiography}[{\includegraphics[width=1in,height=2in,clip,keepaspectratio]{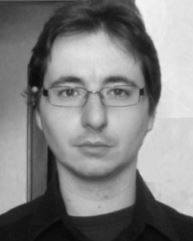}}]{Umberto Montanaro} received the Laurea (M.Sc.)
degree (\textit{cum laude}) in computer science engineering
from the University of Naples Federico II, Naples,
Italy, in 2005, and the Ph.D. degrees in control
engineering and in mechanical engineering from
the University of Naples Federico II, in 2009 and
2016, respectively. From 2010 to 2013, he was a
Research Fellow with the Italian National Research
Council (Istituto Motori). He served as a temporary Lecturer in automation and process control
with the University of Naples Federico II. He is
currently a Lecturer in control systems for automotive engineering with
the Department of Mechanical Engineering Sciences, University of Surrey,
Guildford, U.K. The scientific results he has obtained till now have been
the subject of more than 60 scientific articles published in peer-reviewed
international scientific journals and conferences. His research interests range
from control theory to control application and include: adaptive control, with
special care to model reference adaptive control, control of piecewise affine
systems, control of mechatronics systems, automotive systems, and connected
autonomous vehicles.
\end{IEEEbiography}
\begin{IEEEbiography}[{\includegraphics[width=1in,height=2in,clip,keepaspectratio]{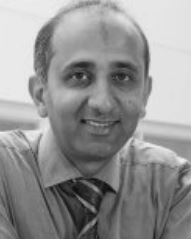}}]
{Saber Fallah} is currently an Associate Professor
with the University of Surrey. He is also the Director
of the Connected and Autonomous Vehicles Lab
(CAV-Lab), Department of Mechanical Engineering
Sciences. Since joining the University of Surrey,
he has been contributing to securing research funding from EPSRC, Innovate UK, EU, KTP, and
industry.\\
He has coauthored a textbook \textit{Electric and Hybrid
Vehicles: Technologies, Modeling and Control-A
Mechatronics Approach} (John Wiley Publishing
Company, 2014). The book addresses the fundamentals of mechatronic design
in hybrid and electric vehicles. His work has contributed to the state-of-the-art
research in the areas of connected autonomous vehicles and advanced driver
assistance systems. So far, his research has produced four patents and more
than 40 peer reviewed publications in high-quality journals and well-known
conferences. He is the Co-Editor of the book of conference proceedings
resulted from the organization of the TAROS 2017 Conference (Springer,
2017).
\end{IEEEbiography}




\end{document}